\documentclass[11pt]{article}
\usepackage{amsmath,amssymb,graphicx,pict2e}
\usepackage{hyperref}
\usepackage{multicol,color,longtable}
\definecolor{darkred}{rgb}{0.65,0.15,0}
\hypersetup{pdfborder={0 0 0},colorlinks=true,urlcolor=blue,citecolor=blue,linkcolor=darkred,linktocpage=true}

\usepackage{bbold}

\usepackage{cite}
\usepackage{array,multirow,stmaryrd}

\usepackage{young}
\usepackage[vcentermath]{youngtab}

\newcommand{\eprint}[1]{{\href{http://arxiv.org/abs/#1}{[\texttt{#1}]}}}
\newcommand{\eprintN}[1]{{\href{http://arxiv.org/abs/#1}{[\texttt{#1 [hep-th]}]}}}
\newcommand{\doi}[2]{{\hypersetup{urlcolor=darkred}\href{http://dx.doi.org/#2}{#1}\hypersetup{urlcolor=blue}}}

\newcommand{\nn}{\nonumber}
\newcommand{\mf}[1]{{\mathfrak{#1}}}
\newcommand{\cN}{\mathcal{N}}

\newcommand{\lb}{\left[}
\newcommand{\rb}{\right]}
\newcommand{\lbb}{\left\llbracket}
\newcommand{\rbb}{\right\rrbracket}
\newcommand{\lc}{\left\{}
\newcommand{\rc}{\right\}}

\newcommand{\ALT}{\textrm{\large{$\wedge$}}}
\newcommand{\SYM}{\textrm{\large{$\vee$}}}
\newcommand{\ad}{\mathrm{ad}\,}

\newcommand{\Minfty}{\textrm{Maxwell}_{\infty}}

\makeatletter

\@addtoreset{equation}{section} \makeatother

\begin{document}

{\flushright {ICCUB-18-017}\\[5mm]}

\begin{center}
{\LARGE \bf Symmetries of M-theory\\[2mm] and free Lie superalgebras}\\[10mm]

\vspace{8mm}
\normalsize
{\large  Joaquim Gomis${}^{1}$, Axel Kleinschmidt${}^{2,3}$ and Jakob Palmkvist${}^4$}

\vspace{10mm}
${}^1${\it Departament de F\'isica Qu\`antica i Astrof\'isica\\ and
Institut de Ci\`encies del Cosmos (ICCUB), Universitat de Barcelona\\ Mart\'i i Franqu\`es , ES-08028 Barcelona, Spain}
\vskip 1 em
${}^2${\it Max-Planck-Institut f\"{u}r Gravitationsphysik (Albert-Einstein-Institut)\\
Am M\"{u}hlenberg 1, DE-14476 Potsdam, Germany}
\vskip 1 em
${}^3${\it International Solvay Institutes\\
ULB-Campus Plaine CP231, BE-1050 Brussels, Belgium}
\vskip 1 em
${}^4${\it Division for Theoretical Physics, Department of Physics, Chalmers University of Technology\\ SE-412 96 Gothenburg, Sweden}

\vspace{20mm}

\hrule

\vspace{10mm}

\begin{tabular}{p{12cm}}
{\small
We study systematically various extensions of the Poincar\'e superalgebra. The most general structure starting from a set of spinorial supercharges $Q_\alpha$ is a free Lie superalgebra that we discuss in detail. 
We explain how this universal extension of the Poincar\'e superalgebra gives rise to many other algebras as quotients, some of which have appeared previously in various places in the literature. In particular, we show how some quotients can be very neatly related to Borcherds superalgebras. The ideas put forward also offer some new angles on exotic branes and extended symmetry structures in M-theory.}
\end{tabular}
\vspace{7mm}
\hrule
\end{center}

\newpage

\setcounter{tocdepth}{2}
\tableofcontents

\vspace{5mm}
\hrule
\vspace{5mm}

\section{Introduction}

Translations $P_a$ in the global Poincar\'e space-time symmetry algebra commute among themselves: $\lb P_a, P_b \rb=0$. This is no longer true when one considers translations in the presence of a constant electro-magnetic background field~\cite{Bacry:1970ye,Bacry:1970du,Schrader:1972zd} and the study of such
deformations of the abelian algebra of translations has led to the investigation of so-called Maxwell algebras\cite{Schrader:1972zd,Beckers:1983gp,Soroka:2004fj,Bonanos:2008ez}. The most general deformation one can allow for (based on Lie algebra cohomology) is
\begin{align}
\lb P_a , P_b \rb = Z_{ab}\,,
\end{align}
where the anti-symmetric generator $Z_{ab}=Z_{[ab]}$ is associated with the constant electro-magnetic background field and transforms as a tensor under Lorentz transformations. In a next step one can analyse the commutator of $Z_{ab}$ with $P_a$ and iterate the procedure~\cite{Bonanos:2008ez}. It was shown in~\cite{Gomis:2017cmt} that the most general structure one obtains in this way is the infinite-dimensional \textit{free} Lie algebra, called $\Minfty$ there, generated by the translations $P_a$. Any free Lie algebra admits many Lie algebra ideals and associated quotient Lie algebras, some of which were studied for $\Minfty$ in~\cite{Gomis:2017cmt}. In particular~\cite{Gomis:2017cmt} identified quotients that could be used in a particle model to describe motion in a general, non-constant, electro-magnetic field in an `unfolded' treatment of the electro-magnetic field strength. This was achieved by introducing additional coordinates for the additional generators contained in the quotient of $\Minfty$. Associating the translations $P_a$ with a coordinate $x^a$ one in this way has an extended space with coordinates $\theta^{ab}$ for $Z_{ab}$ and so on. The dynamical model based on the $\Minfty$ algebra describes the dynamical object together with the background it moves in.

In the present article, we shall extend the picture of the free Lie algebra generalisation of the Poincar\'e algebra to a free Lie superalgebra generalisation of the Poincar\'e superalgebra. The basic building blocks in this construction are the (global) odd supersymmetry generators $Q_\alpha$ (with $\alpha$ an appropriate spinor index). The infinite-dimensional free Lie superalgebra generated by the $Q_\alpha$ has the most general generators in $\lc Q_\alpha, Q_\beta\rc$ and subsequent multi-(anti-)commutators. Again, it admits many non-trivial quotients that we shall discuss. Of particular interest to us will be quotients related to the eleven-dimensional supersymmetry algebra. As has been proved for example in~\cite{vanHolten:1982mx,Townsend:1995gp}, the anti-commutator $\lc Q_\alpha, Q_\beta\rc$ should yield not only the translation generators $P_a$ but also additional non-central terms that serve as sources for branes. These additional terms arise naturally in the free Lie superalgebra as we will show explicitly. Further extensions of the eleven-dimensional supersymmetry algebra were also studied in~\cite{Sezgin:1996cj} (called `M-algebra' there) where non-trivial multi-commutators arise. As we shall describe in more detail below, these again have a natural home in the free Lie superalgebra. The interpretation given in~\cite{Sezgin:1996cj} of some of these generators as superpartners of the bosonic brane charges might give some idea of how to interpret the infinitely many new generators in the free Lie superalgebra.  We will also discuss other possible connections to proposals of infinite-dimensional M-theory symmetries and extended space-times.

Studying the free Lie superalgebra in four space-time dimensions we are also able to make contact with the supersymmetric extension of the Maxwell algebra that was studied in~\cite{Bonanos:2009wy} as it arises as a particular finite-dimensional quotient. In \cite{Bonanos:2009wy}, a dynamical realisation of the supersymmetric Maxwell algebra was found in terms of a  massless superparticle where again additional coordinates are introduced for the new generators and these have interpretations in terms of a Maxwell superfield background. Similar generalisations of the Maxwell algebra for extended supersymmetry have also been investigated~\cite{Kamimura:2011mq} and we shall also exhibit how these $\cN$-extended versions can be obtained from a free Lie superalgebra.

Our paper focusses almost exclusively on the algebraic aspects unifying the various algebras in terms of free Lie superalgebras. We shall not consider particle or string models realising these symmetries but it should be possible to construct these along the lines of~\cite{Green:1989nn}.

An interesting observation, that also opens up connections to other algebraic structures investigated in the literature in connection with potential symmetries of M-theory, is that many quotients of the free Lie superalgebra embed in Kac--Moody~\cite{Kac,West:2001as,Damour:2002cu}, Borcherds~\cite{Borcherds,Henneaux:2010ys,Palmkvist:2012nc} or tensor hierarchy algebras~\cite{Palmkvist:2011vz,Palmkvist:2013vya,Greitz:2013pua,Bossard:2017wxl}. This is due to the definition of these algebras as being constructed as quotients of free algebras. A noteworthy difference is, however, that many of these conjectured algebraic structures in M-theory also have `negative level' symmetry generators that are not present in our free Lie superalgebra constructions. This connection is discussed in detail in section~\ref{sec:Borcherds}.

The paper is structured as follows. We first review and motivate the free Lie algebra construction in relation to the bosonic Maxwell algebra and then introduce the free Lie superalgebra that is the main object of study in this paper. In section~\ref{sec:examples}, we then discuss in detail various examples that 
are related to simple and extended supersymmetry in $D=4$ and simple supersymmetry in $D=11$. We also explain how our construction relates to various algebras in the literature and, potentially, to exotic branes. Section~\ref{sec:Borcherds} contains a discussion of cases where quotients of the free Lie superalgebra can be re-expressed through Borcherds superalgebras and how this in turn can be recast in terms of Kac--Moody algebras. Section~\ref{sec:conc} contains some concluding remarks and speculations.

\section{Review of Maxwell and free Lie (super)algebras}
\label{sec:FLAs}

In this section, we review some basic aspects of the conventional bosonic Maxwell algebra in order to motivate our generalisations first to free Lie algebras~\cite{Gomis:2017cmt} and then to free Lie superalgebras.

\subsection{Maxwell algebra and free Lie algebras}
\label{sec:MaxwellFree}

The starting point for all bosonic constructions here is the Poincar\'e algebra in $D$ space-time dimensions,\footnote{Our anti-symmetrisation convention is of strength one: $M_{[ab]} = \frac12 ( M_{ab} - M_{ba})$. The fundamental Lorentz indices lie in the range $a,b\in \{0,\ldots,D-1\}$. We use the mostly plus convention for the flat Minkowski metric.}
\begin{subequations}
\label{eq:PA}
\begin{align} \label{eq:MMM}
\lb M_{ab} , M_{cd} \rb &= 2\eta_{c[b} M_{a]d} - 2\eta_{d[b} M_{a]c}\,, \\
\lb M_{ab}  , P_c \rb &= 2\eta_{c[b} P_{a]} \,,\\
\label{eq:PP}
\lb P_a , P_b \rb &=0\,.
\end{align}
\end{subequations}
The first line is just the Lorentz algebra $\mf{so}(1,D-1)$ of the anti-symmetric $M_{ab}$ and the second line expresses that the translation generators $P_a$ form a vector representation of the Lorentz algebra. The third line~\eqref{eq:PP} is the standard property of the Poincar\'e algebra that the translations commute. This is the relation that is deformed in the Maxwell algebra and its free Lie algebra generalisation.

As shown in~\cite{Bonanos:2008ez} based on Lie algebra cohomology, the most general deformation of~\eqref{eq:PP} is
\begin{align}
\label{eq:PPZ}
\lb P_a, P_b \rb = Z_{ab}
\end{align}
with a new generator $Z_{ab} = Z_{[ab]}$. This is also the commutator in the free Lie algebra based on the $P_a$ by giving a name to $\lb P_a, P_b \rb$. The algebra spanned by $\{M_{ab}, P_a, Z_{ab}\}$ with trivial commutators $\lb P_a , Z_{bc}\rb = \lb Z_{ab}, Z_{cd}\rb=0$ is the Maxwell algebra introduced 
in~\cite{Schrader:1972zd,Beckers:1983gp,Soroka:2004fj,Bonanos:2008ez}.

From the Lie algebra cohomology perspective, it is possible to deform also these trivial commutation relations. The next most general commutator is~\cite{Bonanos:2008ez,Gomis:2017cmt}
\begin{align}
\label{eq:PPP}
\lb \lb P_a, P_b\rb , P_c \rb = \lb Z_{ab}, P_c\rb =:  Y_{ab,c}\,.
\end{align}
For this to form a consistent Lie algebra, the Jacobi identity has to be obeyed. This implies that the Lorentz tensor $Y_{ab,c}$ has the property $Y_{[ab,c]}=0$. The comma notation for the indices reflects that the tensor also satisfies $Y_{[ab],c} = Y_{ab,c}=- Y_{ba,c}$. As a representation of the symmetric group it has the Young tableau $\raisebox{-0.2\height}{\scalebox{0.5}{\yng(2,1)}}$.\footnote{\label{fn:redSO}We note that this is not an irreducible representation of the Lorentz algebra as one can define the non-trivial trace $\eta^{bc} Y_{ab,c}$.}

This construction can be iterated to obtain all possible multiple commutators of the $P_a$ up to the Jacobi identity (and anti-symmetry). This is by definition the \textit{free Lie algebra} generated by the translation generators $P_a$ and we denote it by $\mf{f}(P)$. It has a natural graded structure given by how many times $P_a$ appears in a multi-commutator. In this grading, we refer to the translation generators $P_a$ as positive level  $\ell=1$, to their commutator $Z_{ab} = \lb P_a, P_b \rb$ as positive level $\ell=2$ and so on. 

The basic generators $P_a$ on level $\ell=1$ transform under the Lorentz algebra (as vectors) and consequently, we can make the generators at any level $\ell$ also transform under the Lorentz algebra. Thus, every level $\ell>0$ of the free Lie algebra $\mf{f}(P)$ consists of Lorentz tensors and we can form the semi-direct sum
\begin{align}
\label{eq:Minfty}
\Minfty = \mf{so}(1,D-1) \oplus \mf{f}(P)
\end{align}
with the $\mf{so}(1,D-1)$ Lorentz generators $M_{ab}$. This semi-direct sum is the infinite generalisation $\Minfty$ of the Poincar\'e and Maxwell algebra introduced in~\cite{Gomis:2017cmt}. We will refer to the algebra $\mf{so}(1,D-1)$ as level $\ell=0$.

At this point it is important to remark that the construction of a free Lie algebra $\mf{f}(P)$ only requires a set of $D$ generators $P_a$, but that the transformation property of the $P_a$ under level $\ell=0$ is not something that is fixed by the free Lie algebra construction. In fact, we could also view the generators $P_a$ as transforming in the fundamental of $\mf{gl}(D)$ or any other Lie algebra $\mf{g}_0$ with a $D$-dimensional representation. This representation need not even be irreducible. This level $\ell=0$ algebra $\mf{g}_0$ will then act on all the positive levels 
as well and we could arrange the higher levels in irreducible representations of $\mf{g}_0$. Choosing for instance $\mf{g}_0=\mf{gl}(D)$, one obtains tensors of $\mf{gl}(D)$ at higher levels. As $\mf{gl}(D)$ does not have an invariant trace, the generator $Y_{ab,c}$ at $\ell=3$ then is irreducible while it is reducible under $\mf{so}(1,D-1)$, see footnote~\ref{fn:redSO}. For yet different choices of $\mf{g}_0$ acting on $P_a$ each level $\ell$ arranges differently into irreducible representations of $\mf{g}_0$; the total number of generators on each level of course does not change as this is determined by $\mf{f}(P)$ which is defined independently of the choice of $\mf{g}_0$.\footnote{Other interesting choices of $\mf{g}_0$ relate to non-relativistic systems with a Galiliean symmetry or very special relativity~\cite{Cohen:2006ky} where only the $\text{Sim}(D-2)$ algebra with generators $M_{+-}$, $M_{+i}$ and $M_{ij}$ in a light-cone basis $\{ x^\pm, x^i\}$ is kept~\cite{Gibbons:2009me}.}

The structure of a free Lie algebra $\mf{f}(P)$ can be computed from a generating function~\cite{Cederwall:2015oua} or explicitly using the Witt formula~\cite{Witt} that was reviewed in~\cite{Gomis:2017cmt}. 
The generating function identity governing the free Lie algebra is
\begin{align}
\label{eq:genB}
\bigotimes_{\ell=1}^\infty \left[\bigoplus_{k=0}^\infty (-1)^k t^{k\ell} \ALT^k T_\ell \right]= 1-tT_1 \,.
\end{align}
In this formula, $t$ is a formal parameter and $T_\ell$ denotes the vector space of generators at level $\ell$, such that $T_1=\left\langle P_a\,\middle|\, a=0,\ldots,D-1 \right\rangle$ and $T_1$ plays a special role since it fully determines $\mf{f}(P)$. The identity \eqref{eq:genB} can be derived from the denominator formula for a Borcherds algebra. Expanding out the generating identity yields the following recursive identities at the first few powers of $t$:
\begin{subequations}
\label{eq:recB}
\begin{align}
T_1 &= \left\langle P_a \right\rangle\,,\\
T_2 &= \ALT^2 T_1\,,\\
T_3 &= T_2 \otimes T_1 \ominus \ALT^3 T_1\,,\\
T_4 &= (T_3 \otimes T_1 \oplus \ALT^2 T_2) \ominus T_2 \otimes \ALT^2 T_1 \oplus \ALT^4 T_1\,,\\ 
T_5 &= (T_4 \otimes T_1 \oplus T_3 \otimes T_2) \ominus (T_3 \otimes \ALT^2 T_1 \oplus \ALT^2 T_2\otimes T_1)\oplus T_2 \otimes \ALT^3T_1 \ominus \ALT^5T_1\,,\\
&\cdots \nn
\end{align}
\end{subequations}
The notation here is such that $\ALT^k R$ denotes the $k$-th anti-symmetric tensor power of a representation.
With $\ominus$ we denote the removal of representations from the module. The lowest levels are simple to understand as representing the anti-symmetry of the Lie bracket and the Jacobi identity of the free Lie algebra: $T_2$ consists of all commutators of the $P_a$ in $T_1$ and therefore is the anti-symmetric product. 
The vector space $T_3$ is spanned by all commutators of $P_a$ in $T_1$ with the elements in $T_2$, as in~\eqref{eq:PPP}, leading to the tensor product. The totally anti-symmetric Jacobi identity has to be satisfied by the triple commutators at $\ell=3$ and so one has to remove $\ALT^3 T_1$ from the tensor product.

The tensor products here can be thought of as either just vector space operations or --and this will be more useful to us-- as tensor products of representations of $\mf{g}_0$ which is the level $\ell=0$ algebra under which the $P_a$ transform. The choice in~\eqref{eq:Minfty} is $\mf{g}_0= \mf{so}(1,D-1)$ and the tensor products would be those of Lorentz algebra representations.

Free Lie algebras admit many ideals and associated quotient Lie algebras. Several examples were discussed in~\cite{Gomis:2017cmt} related to particle models in electro-magnetic backgrounds and previous examples in the literature.
Any set of elements $X$ in $\mf{f}(P)$ generates an ideal,
spanned by all
multi-commutators obtained by acting on the elements
$X$ with the basic generators $P_a$, that is, the ideal is spanned by $[P,[P,\ldots,[P,X]]]$. If the set $X$ consists of only one element,
then this is called a principal ideal. One can also consider ideals that are generated by a set which is a subspace of $\mf{f}(P)$. In fact, any ideal generated by a set of elements $X$ is the same as the ideal generated by the subspace $\langle X \rangle$ spanned by these elements.

The subspace $\langle X \rangle$ of 
$\mf{f}(P)$ may be equal to $T_{k+1}$ for some fixed finite level $k$. Then the ideal consists of all elements in $\mf{f}(P)$ with more than $k$ factors of $P_a$ in the multi-commutator or, equivalently, the direct sum of all $T_\ell$ with $\ell>k$, and the quotient is a finite-dimensional truncation of $\mf{f}(P)$ including
only generators up to level $k$. For example, the truncation to $k=1$ recovers the standard Poincar\'e algebra while the truncation to $k=2$ gives the Maxwell algebra with $P_a$ and $Z_{ab}$ in the quotient.

More generally, the vector space $\langle X \rangle$ may be a subspace of $T_{k+1}$,
and then the quotient may be infinite-dimensional.
The Serre ideals appearing in the construction of
Borcherds or Kac--Moody algebras discussed in section~\ref{sec:Borcherds} are generated in this way.
Even more generally, the vector space $\langle X \rangle$ may be decomposed into a direct sum $X=s_1\oplus\ldots\oplus s_n$ of finitely or infinitely many
subspaces, or representations of $\mf{g}_0$. These representations $s_i$ need not all be on the same level. Again, the quotient may be infinite-dimensional.
An example of such an infinite-dimensional quotient was discussed in~\cite{Gomis:2017cmt} where a realisation in terms of particles in general `unfolded' electro-magnetic backgrounds was achieved.

As an example of a free Lie algebra that has not appeared in the literature before to the best of our knowledge, we consider the case of $D=3$ translations $P_a$ transforming under the Lorentz algebra $\mf{g}_0= \mf{so}(1,2)$ whose complexification is of Cartan type $A_1$. The corresponding lowest levels are listed in table~\ref{tab:3D}. Using standard mathematical Dynkin labels, the translation generators are denoted by $[2]$.

\begin{table}[t!]
\centering
\renewcommand{\arraystretch}{1.3}
\begin{tabular}{c|c|c}
$T_\ell$ & representation & generator\\
\hline\hline
$T_1$ & $[2]$ & $P_a$\\\hline
$T_2$ & $[2]$ & $Z_{ab}= \epsilon_{abc} Z^c$\\\hline
$T_3$ & $[2]\oplus [4] $ & $Y_{ab,c}= \epsilon_{abd} \tilde{Y}^d{}_c$\\\hline
\multirow{2}{10mm}{\centering$T_4$} & $[2]\oplus [4] \oplus [6]$ & $S^1_{ab,c,d}$\\
& $[2$] &  $S^2_a$\\\hline
$\vdots$& $\vdots$ & $\vdots$
\end{tabular}
\caption{\label{tab:3D}\it The free Lie algebra $\mf{f}(P)$ in $D=3$ space-time dimensions. Representations are labelled by the complexified Lorentz algebra of type $A_1$ while the generators represent the associated $\mf{so}(1,2)$ tensors. For mixed symmetry, these are reducible in the way described in footnote~\ref{fn:redSO}.}
\end{table}

The generators listed in table~\ref{tab:3D} can also be obtained by dimensional reduction from the general result in~\cite{Gomis:2017cmt}. The three-dimensional case is of particular interest since one can find quotients of the $\Minfty$ algebra that admit an invariant and non-degenerate bilinear form
that then can be used to construct three-dimensional Chern--Simons theory~\cite{Salgado:2014jka}. Since three-dimensional Chern--Simons theory is related to gravity, this leads to a Maxwell version of gravity ~\cite{Salgado:2014jka,Salgado:2014qqa,Hoseinzadeh:2014bla,Caroca:2017izc} with a supersymmetric version studied in~\cite{Concha:2018jxx} and  also the non-relativistic limit has been studied in this case~\cite{Aviles:2018jzw}. A four-dimensional gravity system with Maxwell symmetry has been investigated in~\cite{deAzcarraga:2010sw}.

\subsection{Free Lie superalgebras}
\label{sec:FLSA}

The bosonic free Lie algebra $\mf{f}(P)$ serves as a very universal object to construct kinematical algebras relevant in various contexts. It is natural to consider the corresponding Lie superalgebra and in this section we give some general properties of its construction. Explicit examples will be discussed in the next section.

We shall consider the free Lie superalgebra generated by odd supertranslations $Q_\alpha$ where $\alpha$ labels the 
supersymmetry generators and we here think of them as spinorial generators. The exact range depends on the number of space-time dimensions and on whether one consider simple or extended supersymmetry. The general discussion of this section is independent of these details. 

The free Lie superalgebra $\mf{f}(Q)$ is given by all possible graded multi-commutators of the basic $Q_\alpha$ and subjected to the graded Jacobi identity
\begin{align}
\label{eq:SJ}
\lb \lc Q_\alpha, Q_\beta\rc , Q_\gamma\rb + \lb \lc Q_\beta, Q_\gamma\rc , Q_\alpha\rb + \lb \lc Q_\gamma, Q_\alpha\rc , Q_\beta\rb =0\,.
\end{align}
 `Graded' here refers to the $\mathbb{Z}_2$-grading of superalgebras; the $Q_\alpha$ are by definition odd while their graded commutator $\lc Q_\alpha, Q_\beta\rc$ is even. One can again define a level $\ell$ on $\mf{f}(Q)$ by counting the number of $Q_\alpha$ that occur in a general graded multi-commutator. For odd $\ell$ one obtains generators that are also odd in the $\mathbb{Z}_2$ sense while even $\ell$ gives $\mathbb{Z}_2$ even generators such that the two gradings are consistent. 

Similar to the bosonic case one can describe the structure of the free Lie superalgebra in terms of a generating function. The generalisation of~\eqref{eq:genB} is given by
\begin{align}
\label{eq:genF}
\bigotimes_{\ell\,\text{odd}} \left[ \bigoplus_{k=0}^\infty (-1)^k t^{k \ell} \SYM^k R_\ell \right] \otimes \bigotimes_{\ell\, \text{even}} \left[ \bigoplus_{k=0}^\infty (-1)^k t^{k \ell} \ALT^k R_\ell \right] =1 - t R_1\,.
\end{align}
Here, $t$ is again a formal parameter and the $\mathbb{Z}_2$ nature of the generators introduces some signs.
 $R_1=\langle Q_\alpha \rangle$ denotes the generators on level $\ell=1$ and we denote the generators on level $\ell$ by $R_\ell$. 
 Expanding out the generating identity one obtains for the first few levels
\begin{subequations}
\label{eq:recF0}
\begin{align}
R_1 &= \left\langle Q_\alpha \right\rangle\,,\\
R_2 &= \SYM^2 R_1\,,\\
R_3 &= R_2 \otimes R_1 \ominus \SYM^3 R_1\,,\\
R_4 &= (R_3 \otimes R_1 \oplus \ALT^2 R_2) \ominus  R_2\otimes\SYM^2 R_1 \oplus \SYM^4 R_1\,,\\
R_5 &= (R_4 \otimes R_1 \oplus R_3 \otimes R_2) \ominus (R_3 \otimes \SYM^2 R_1 \oplus \ALT^2 R_2\otimes R_1)\oplus R_2 \otimes \SYM^3R_1 \ominus \SYM^5R_1\,,\\
&\cdots \nn
\end{align}
\end{subequations}
Compared to (\ref{eq:recB}), the anti-symmetric tensor powers $\ALT^k$ are interchanged for symmetric ones, denoted $\SYM^k$, in a few places.
For instance $R_2$ is given by all \textit{symmetric} products $\lc Q_\alpha, Q_\beta\rc$ of the $Q_\alpha$ and the Jacobi identity relevant for $R_3$ is totally symmetric as well.

We can again consider the case when the $Q_\alpha$ transform under some level $\ell=0$ algebra $\mf{g}_0$. This could be the Lorentz algebra $\mf{so}(1,D-1)$ in $D$ dimensions or could also contain possible R-symmetries. We shall consider many examples below. Given a $\mf{g}_0$, the generators $R_\ell$ can be arranged into representations of $\mf{g}_0$ by treating the tensor products as products of $\mf{g}_0$ modules.

Just as free Lie algebras, free Lie superalgebras admit many quotients and one can again distinguish the case of finite-dimensional and infinite-dimensional quotients. Examples of extensions of the standard supersymmetry algebra that have been discussed in the literature are finite-dimensional quotients and we shall illustrate this in examples below. Moreover, there are cases where these quotients can be described conveniently in terms of Borcherds superalgebras and Dynkin diagrams as we discuss in section~\ref{sec:Borcherds}.

\section{Instances of free Lie superalgebras} 
\label{sec:examples}

In this section, we apply the algorithm~\eqref{eq:recF0} 
and construct the free Lie superalgebra generated by $Q_\alpha$
in various cases where $\mf{g}_0$ is either the Lorentz algebra in $D$ dimensions, or the direct sum of the Lorentz algebra and an R-symmetry algebra
$\mf{su}(\cN)$.

\subsection{$D=4$ and $\cN=1$ supersymmetry} \label{sec:D4N1}

We begin with the free Lie superalgebra generated by supersymmetry generators $Q_\alpha$ and then compare the results to various extensions of
the Poincar\'e superalgebras that have appeared in the literature.

\subsubsection{Free Lie superalgebra}

Following the philosophy outlined in the previous section, we now construct the \textit{free Lie superalgebra} $\mf{f}(Q)$ generated by the $Q_\alpha$ in $D=4$. The spinor index $\alpha=1,\ldots,4$ here labels the four independent components of a Majorana spinor. Taking $\mf{g}_0=\mf{so}(1,3)$, the transformation of $R_1=\langle Q_\alpha\rangle$ is given by
\begin{align}
\label{eq:MQ}
\lb M_{ab}, Q_\alpha \rb &= 
\frac12  Q_\beta (\Gamma_{ab})^{\beta}{}_\alpha \,.
\end{align}
The real gamma matrices here satisfy $\lc \Gamma_a, \Gamma_b\rc = 2\eta_{ab}$. Gamma matrices with multiple indices are defined by $\Gamma^{ab} = \Gamma^{[a} \Gamma^{b]}$ etc. The gamma matrices that we use are given explicitly in Appendix~\ref{gammapp}. We use the signature $\eta=(-++\,+)$ and below we shall also encounter the symmetric combinations   $(C\Gamma^a)^T = C\Gamma^a$ and $(C\Gamma^{ab})^T = C\Gamma^{ab}$ involving the  charge conjugation matrix $C$.

The free Lie superalgebra $\mf{f}(Q)$ is obtained by taking all possible (graded) commutators of the $Q_\alpha$ obeying only graded anti-symmetry and the Jacobi identity. Following the algorithm~\eqref{eq:recF0} this leads to table~\ref{tab:SM4}. In this table we have labelled the representations of the Lorentz algebra 
$\mf{g}_0=\mf{so}(1,3)$ in terms of Dynkin labels of 
the corresponding complex Lie algebra $A_1 \oplus A_1$, 
such that a Majorana spinor decomposes as
$Q_\alpha \leftrightarrow [1,0]\oplus [0,1]$, the familiar decomposition into a left-handed and right-handed spinor in $D=4$.
Keeping this in mind, 
we can use the LiE software~\cite{LiE} to compute
the tensor product decompositions of $A_1 \oplus A_1$. Generally, a complex representation of $A_1\oplus A_1$ gives a real representation of $\mf{so}(1,3)$ if the representations of the two $A_1$ are balanced, i.e., either of the symmetric form $[p_1,p_2] \oplus [p_2,p_1]$ of directly $[p,p]$.

\begin{table}[t!]
\centering
\renewcommand{\arraystretch}{1.3}
\begin{tabular}{c|c|c}
$R_\ell$ & representation & generator\\
\hline\hline
$R_1$ & $[1,0]\oplus [0,1]$ & $Q_\alpha$\\\hline
\multirow{2}{10mm}{\centering$R_2$} & $[1,1]$ & $P_a$\\
& $[2,0]\oplus [0,2]$ & $P_{ab}=P_{[ab]}$\\\hline
\multirow{3}{10mm}{\centering$R_3$} & $[1,0]\oplus [0,1]$ & $\Sigma_\alpha$\\
& $[1,0]\oplus [0,1]$ & $\tilde{\Sigma}_\alpha$\\
& $[1,2]\oplus [2,1]$ & $\Sigma_{a\, \alpha}$\\\hline
\multirow{6}{10mm}{\centering$R_4$} & $[2,0]\oplus [0,2]$ & $Z_{ab}=Z_{[ab]}$\\
& $[0,0]$ & $B$\\
& $3\times [1,1]$ & vectorial\\
& $[3,1]\oplus [1,3]\oplus[1,1]$ & $\raisebox{-0.2\height}{\scalebox{0.5}{\yng(2,1)}}$\\
& $[2,2]\oplus [0,2]\oplus[2,0]$ & $\raisebox{-0.2\height}{\scalebox{0.5}{\yng(2,1,1)}}$\\
& $[2,0]\oplus [0,2]$ & two-form\\\hline
$\vdots$& $\vdots$ & $\vdots$
\end{tabular}
\caption{\label{tab:SM4}\it The free Lie superalgebra $\mf{f}(Q)$ in $D=4$ space-time dimensions for a single four-component spinor generator $Q_\alpha$. We have grouped representations together that form a nice representation of $\mf{so}(1,3)$. All tensor-spinors are gamma traceless, but we retain the tensor traces, meaning that for example the first non-trivial Young tableau in $R_4$ represents a tensor structure $Y_{ab,c}$ where one can still take the Lorentz trace as discussed in footnote~\ref{fn:redSO}; this corresponds to the $[1,1]$ representation listed there. Some representations occur with a non-trivial multiplicity and we have not given names to generators that do not appear elsewhere in this article.}
\end{table}

Let us also give the commutation relations for the free Lie superalgebra $\mf{f}(Q)$ at lowest levels, using the names for the generators introduced in table~\ref{tab:SM4}. The first non-trivial commutator is that of the supercharges $Q_\alpha$ that we define to be
\begin{align}
\lc Q_\alpha, Q_\beta\rc = (C\Gamma^a)_{\alpha\beta} P_a + \frac12 (C\Gamma^{ab})_{\alpha\beta} P_{ab}\,,  \label{eq:QQfree}
\end{align}
where the term $P_{ab}=P_{[ab]}$ has already appeared in~\cite{Achucarro:1988qb,Ferrara:1997tx,Chryssomalakos:1999xd}. It does not appear in the standard $\cN=1$ supersymmetry algebra and it is not central as it does transform non-trivially under the Lorentz algebra. It can be interpreted as a source for the supersymmetric membrane in $D=4$~\cite{Achucarro:1988qb} in a way similar to the complete $\lc Q,Q \rc$ commutators in eleven dimensions~\cite{Townsend:1995gp}. We shall come back to this interpretation below. We also note that~\eqref{eq:QQfree} implies
\begin{align}
P_a = \frac14 (\Gamma_a C^{-1})^{\alpha\beta} \lc Q_\alpha, Q_\beta\rc\,,\quad
P_{ab} = -\frac14 (\Gamma_{ab} C^{-1})^{\alpha\beta} \lc Q_\alpha, Q_\beta\rc\,.
\end{align}

In the free Lie superalgebra, one can then form the triple commutators $\lb \lc Q_\alpha, Q_\beta\rc,Q_\gamma\rb$ at level $\ell=3$ that have to be subjected to the Jacobi identity~\eqref{eq:SJ}. The resulting most general expression can be split up into the commutators of $Q_\alpha$ with $P_a$ and $P_{ab}$ as
\begin{subequations} \label{eq:QPfree}
\begin{align}
\lb  Q_\alpha ,P_a \rb &= \Sigma_\beta (\Gamma_a)^\beta{}_\alpha +  \Sigma_{a\, \alpha}\,,\\
\label{eq:QPab}
\lb  Q_\alpha ,P_{ab} \rb &= \tilde{\Sigma}_{\beta}(\Gamma_{ab})^\beta{}_\alpha +\frac43  \Sigma_{[a\, \beta} (\Gamma_{b]})^\beta{}_\alpha\,.
\end{align}
\end{subequations}
The anti-symmetrisation only refers to the vector indices $a,b$. As mentioned in the caption of table~\ref{tab:SM4}, the tensor spinors we use are gamma traceless which means that
\begin{align}
\Sigma_{a\, \alpha} (\Gamma^a)^\alpha{}_\beta = 0\,,
\end{align}
and accordingly they span a subspace of dimension $4\times 4-4=12$.
We also note that the Jacobi identities can imply that a naive counting of generators in commutation relations does not always work. For instance,
in~\eqref{eq:QPfree}
the left-hand sides have superficially $4\times 4+4\times 6=40$ elements while the right-hand side has $4+12+4=20$ elements and the discrepancy is due to the Jacobi identity, meaning that there are
$20$ linear combinations of the $40$ possible commutators that vanish due to the Jacobi identity. The consistency of the free Lie superalgebra commutation relations with the Jacobi identity requires the non-trivial gamma matrix identities~\eqref{eq:cyclics}. We shall see below how a quotient relates to the Maxwell superalgebra introduced in~\cite{Bonanos:2009wy}.

Continuing to level $\ell=4$ we have to commute the level $\ell=3$ generators with the basic $Q_\alpha$ on level $\ell=1$. We shall not give the full commutations at this level but restrict ourselves to defining a part that is relevant to the comparison below,
\begin{align}
\label{eq:QS}
\lc Q_\alpha, \Sigma_\beta \rc&= 
\frac14 (C\Gamma^{ab})_{\alpha\beta} Z_{ab}  +(C\Gamma_5)_{\alpha\beta}B+\ldots
\end{align}
such that 
\begin{align} \label{PP=Z}
\lb P_a,P_b\rb=Z_{ab}+\ldots\,,
\end{align}
showing the relation between the free Lie superalgebra and the bosonic free Lie algebra. 
In the two equations above the dots denote additional generators that are present and crucial in the free Lie superalgebra. Their tensor type has been listed in table~\ref{tab:SM4}. For the comparison in the next section we do not require the precise form of these terms.

\subsubsection{Comparison to Maxwell superalgebra in the literature}

Let us now leave the free Lie superalgebra $\mf{f}(Q)$ for the moment and go back to the 
Poincar\'e superalgebra in $D=4$ (with $\cN=1$ supersymmetry):
\begin{subequations}
\begin{align} 
\label{eq:MP}
\lb M_{ab},  P_c \rb &= 
2\eta_{c[b} P_{a]}\,,\\
\lb M_{ab}, Q_\alpha \rb &= 
\frac12  Q_\beta (\Gamma_{ab})^{\beta}{}_\alpha \,.\\
\label{eq:QQ}
\lc Q_\alpha, Q_\beta \rc &=  (C\Gamma^a)_{\alpha\beta} P_a\,,\\
\label{eq:PQ}
\lb P_a,Q_\alpha \rb&=0\,,\\
\label{eq:PP2}
\lb P_a, P_b \rb &=0\,.
\end{align}
\end{subequations}

The Poincar\'e superalgebra has non-trivial cohomology and it can be extended as was studied in~\cite{Bonanos:2009wy}.\footnote{We note that compared to~\cite{Bonanos:2009wy} we have slightly changed conventions by removing some factors of $i$ and rescaling some generators.}
It was shown there that one can in particular deform the commutators~\eqref{eq:PQ} and~\eqref{eq:PP2} through the introduction of a new (non-central) Majorana spinor generator $\Sigma_\alpha$ by letting\footnote{A similar extension of the supersymmetry algebra in $D=10$ dimensions had been introduced previously in order to write the Wess--Zumino term of the
Green--Schwarz string as an invariant term~\cite{Green:1989nn}. The algebra of~\cite{Green:1989nn} has also been extended to study supersymmetric $p$-brane models and their $\kappa$-symmetry~\cite{Bergshoeff:1989ax,Bergshoeff:1995hm}.}
\begin{align} 
\label{PQ=Sigma}
\lb Q_\alpha , P_a \rb&= \Sigma_\beta (\Gamma_a)^\beta{}_\alpha \,.
\end{align}
In order to obtain a minimal $\cN=1$ supersymmetric extension of the Maxwell algebra that includes the generators $\{M_{ab}, P_a, Z_{ab}\}$ where according to~\eqref{eq:PPZ}
\begin{align}
\lb P_a, P_b \rb =Z_{ab}
\end{align}
one also has to impose by the Jacobi identity that
\begin{align}
\{Q_\alpha,\Sigma_\beta\}=\frac14 (C\Gamma^{ab})_{\alpha\beta}Z_{ab}+(C\Gamma_5)_{\alpha\beta}B\,.
\end{align}
In the relation above we have also included a generator $B$ that was introduced in~\cite{Bonanos:2009wy} to pair with the bosonic Lorentz scalar chirality operator $B_5$ satisfying $\lb B_5, Q_\alpha \rb = Q_\beta (\Gamma_5)^\beta{}_\alpha$. From the commutation relations above one then deduces $\lb B_5 , P_a \rb=0$ and $\lb B_5 , \Sigma_\alpha\rb = - \Sigma_\beta (\Gamma_5)^\beta{}_\alpha$. The total algebra considered in~\cite{Bonanos:2009wy} then consists of
\begin{align}
\label{eq:BGKL}
\underbrace{ M_{ab}, B_5}_{\mf{g}_0}, \underbrace{Q_\alpha}_{\ell=1}, \underbrace{P_a}_{\ell=2}, \underbrace{\Sigma_\alpha}_{\ell=3}, \underbrace{Z_{ab}, B}_{\ell=4}
\end{align}
We thus see that the generators of~\cite{Bonanos:2009wy} form a subset of $\mf{g}_0$ together with the free Lie superalgebra $\mf{f}(Q)$ defined above and is consistent with the commutation relations. It is in fact a quotient of the free Lie superalgebra and the quotient can, moreover, be described in terms of a Borcherds superalgebra as we shall explain in more detail in section~\ref{sec:Borcherds}. We also note that the algebra~\eqref{eq:BGKL} admits an invariant Casimir of the form
\begin{align}
\label{eq:Cas}
C_2 =  \frac12 P_a P^a +Q_\alpha C^{\alpha\beta} \Sigma_{\beta}  -  \frac12 M_{ab} Z^{ab} + B_5 B  \,.
\end{align}
The quadratic Casimir~\eqref{eq:Cas} pairs generators on levels $\ell$ and $4-\ell$; extending this beyond the generators in~\eqref{eq:BGKL} would require generators on negative levels that pair with generators on level $\ell>4$. Such a structure is provided by a tensor hierarchy algebra~\cite{Palmkvist:2011vz,Palmkvist:2013vya,Greitz:2013pua,Bossard:2017wxl}, yet another algebra that can be defined from the Poincar\'e superalgebra.
In such an extension, the Casimir would also be made more manifestly graded symmetric by letting $B_5B \to \tfrac12 (B_5 B + BB_5)$, $Q_\alpha C^{\alpha\beta}\Sigma_\beta\to \frac12 (Q_\alpha C^{\alpha\beta}\Sigma_\beta  + \Sigma_\alpha C^{\alpha\beta}Q_\beta)$, etc.; in the algebra~\eqref{eq:BGKL} this is not necessary as these generators commute up to central terms.

\subsection{$D=4$ and extended supersymmetry}
\label{sec:D4ext}

The construction of a free Lie superalgebra $\mf{f}(Q)$ described in section~\ref{sec:FLSA} allows for the $Q$ to be odd generators transforming under some level $\ell=0$ algebra $\mf{g}_0$. In the previous section, we considered the $D=4$ Lorentz algebra $\mf{g}_0 = \mf{so}(1,3)$ (possibly extended by a chirality operator). In this section, we shall consider the case of extended supersymmetry where there is a non-trivial R-symmetry acting on the supersymmetry generators. That is, we consider the case $\mf{g}_0 = \mf{so}(1,3)\oplus \mf{su}(\cN)$ for $\cN$-extended supersymmetry. The corresponding basic generators will be denoted by $Q_\alpha^I$ where $\alpha=1,\ldots,4$ is the Lorentz spinor index while $I=1,\ldots,\cN$ is the R-symmetry index.

\subsubsection{$D=4$ and $\cN=2$ supersymmetry}

Complexifying the algebra as before we are therefore considering $A_1\oplus A_1 \oplus A_1$ where each $A_1$ denotes a complexified $\mf{su}(2)$. Then the level $\ell=1$ generators $Q_\alpha^I$ have representation labels
\begin{align}
\label{eq:N2}
R_1 = [1,0;1] \oplus [0,1;1]\,,
\end{align}
so that the last label is the R-symmetry label, separated with a semicolon. The case $\cN=2$ is special as the R-symmetry representation $[1]$ can be treated as a real representation using $\epsilon^{IJ}$  as a complex structure. This is no longer the case for $\cN>2$ and we treat this case on its own below.  Applying the algorithm~\eqref{eq:recF0} to the $\cN=2$ generator~\eqref{eq:N2} leads to table~\ref{tab:SM42}, where we do not give the result for $R_\ell$ with $\ell>2$ due to a proliferation of terms.

\begin{table}[t!]
\centering
\renewcommand{\arraystretch}{1.3}
\begin{tabular}{c|c|c}
$R_\ell$ & representation & generator\\
\hline\hline
$R_1$ & $[1,0;1]\oplus [0,1;1]$ & $Q_\alpha^I$\\\hline
\multirow{4}{10mm}{\centering$R_2$} & $[1,1;0]$ & $P_a$\\
& $2\times [0,0;0]$ & $P$, $P_5$\\
& $[1,1;2]$ & $P_a^{IJ}=P_a^{(IJ)}$\\
& $[2,0;2]\oplus [0,2;2]$ & $P_{ab}^{IJ}=P_{ab}^{(IJ)}=P_{[ab]}^{IJ}$\\\hline
$\vdots$& $\vdots$ & $\vdots$
\end{tabular}
\caption{\label{tab:SM42}\it The free Lie superalgebra $\mf{f}(Q)$ in $D=4$ space-time dimensions for extended $\cN=2$ supersymmetry. The $\mf{su}(2)$ R-symmetry representation is given by the last Dynkin label while the first two specify the $A_1 \oplus A_1$ representation. The fundamental R-symmetry index is $I=1,2$.}
\end{table}

The commutation relation in the free Lie superalgebra representing the first two levels is 
\begin{align}
\lc Q_\alpha^I, Q_\beta^J \rc &= \epsilon^{IJ} (C\Gamma_5\Gamma^a)_{\alpha\beta} P_a + \epsilon^{IJ}C_{\alpha \beta} P + \epsilon^{IJ} (C\Gamma_5)_{\alpha\beta} P_5 \nn\\
&\quad + (C\Gamma^a)_{\alpha\beta} P_a^{IJ} + \frac12 (C\Gamma_5 \Gamma^{ab})_{\alpha\beta} P_{ab}^{IJ}\,.
\end{align}
The first line corresponds to the central extension of the $\cN=2$ Poincar\'e superalgebra by an electric and magnetic charge. 
The second line introduces additional string and membrane charges~\cite{Ferrara:1997tx,Townsend:1995gp}. 

The calculation can be carried out to higher levels and will contain the hallmark relation $\lb P_a, P_b\rb=Z_{ab}$ of Maxwell algebras. An $\cN$-extended supersymmetric version of the Maxwell algebra in $D=4$ was introduced in~\cite{Kamimura:2011mq} from a contraction of a superconformal algebra. Their algebra has an R-symmetry of $\mf{so}$-type rather than the $\mf{su}(\cN)$ that we have assumed here and therefore the resulting commutation relations are quite different, allowing in particular a raising and lowering of R-symmetry indices.

\subsubsection{$D=4$ and $\cN>2$ supersymmetry}

One can similarly consider $\cN$-extended supersymmetry in $D=4$ dimension for $\cN>2$ by letting the generating elements of the free Lie superalgebra be $Q_\alpha^I$ with $I$ a fundamental index of $\mf{su}(\cN)$. In order to write an anti-commutator of the elementary $Q_\alpha^I$ that yields standard translation generators $P_a$ it is necessary to use 2-component Weyl spinors that we denote by $Q_{A}^I$ and $\bar{Q}_{\dot{A}\ I}$ (with $A=1,2$ and $\dot{A}=\dot{1},\dot{2}$). The important point here is that the two Weyl spinors transform in conjugate (fundamental) representations of $\mf{su}(\cN)$. For $\cN=2$ we could use the invariant $\epsilon^{IJ}$ to relate the two representations, but for $\cN>2$ there is no corresponding invariant tensor to achieve this. The complexification of $\mf{g}_0 = \mf{so}(1,3) \oplus \mf{su}(\cN)$ is of type $A_1\oplus A_1\oplus A_{\cN-1}$ and the corresponding complex representation is
\begin{align}
R_1=[1,0;1,0,\ldots,0] \oplus [0,1;0,\ldots,0,1]\,,
\end{align}
illustrating the that the two Weyl spinors transform in conjugate representations under $A_{\cN-1}$. This space has complex dimension $\dim_{\mathbb{C}} (R_1) = 4\cN$ which is twice the number of real supertranslations of $\cN$-extended supersymmetry. A real slice through the complex representation is picked by a reality condition of the form
\begin{align}
\label{eq:Qconj}
\bar{Q}_{\dot{A}\ I} = (Q_A^I)^*\,.
\end{align}
We note that the conjugation affects the R-symmetry index and spinor index at the same and the R-symmetry transformations in a real basis will therefore also affect the spinor index. This happens for instance for $\cN=8$ supergravity~\cite{Cremmer:1979up,deWit:1982bul}. Writing the $\mf{su}(\cN)$ in terms of its real $\mf{so}(\cN)$ subalgebra one can make the transition between the complex Weyl and a real Majorana basis manifest, but at the sake of giving up manifest $\mf{su}(\cN)$ invariance. We shall not carry out this rewriting here but instead work with the (complex) Weyl spinors.

In terms of the Pauli matrices $\sigma^a_{A\dot{A}}$ one then obtains the (anti-)commutation relations
\begin{subequations}
\begin{align}
\lc Q^I_A,\bar{Q}_{\dot A\ J}\rc &= 2\delta^I_J (\sigma^a)_{A\dot{A}} P_a 
  + (\sigma^a)_{A\dot A} \tilde{P}_a^I{}_J\,\\
\lc Q^I_A,Q^J_B\rc &= P_{AB}^{IJ} + \epsilon_{AB} P^{IJ}\,\\
\lc \bar{Q}_{\dot{A}\ I}, \bar{Q}_{\dot{B}\ J}\rc &= \bar{P}_{\dot{A}\dot{B}\ IJ} + \epsilon_{\dot{A}\dot{B}} \bar{P}_{IJ}\,,
\end{align}
\end{subequations}
where 
\begin{align}
\tilde{P}_a^I{}_I&=0\,, & P_{AB}^{IJ} &= P_{(AB)}^{IJ}=P_{AB}^{(IJ)}\,, & P^{IJ}&= P^{[IJ]}\,,
\end{align}
with similar relations for the conjugate generators. The generators $P_{AB}^{IJ}$ and $\bar{P}_{\dot{A}\dot{B}\ IJ}$ with symmetric pairs of spinor indices combine into an antisymmetric two-form under the Lorentz group $\mf{so}(1,3)$ but this obscures their $\mf{su}(\cN)$ properties in a way similar to~\eqref{eq:Qconj}.

\subsection{$D=11$ and $\cN=1$ supersymmetry}

The last example we discuss in some detail is given by minimal supersymmetry in $D=11$ space-time dimensions as this is the case relevant to M-theory.

\subsubsection{Free Lie superalgebra}
\label{sec:FLA11}

In this case the $Q_\alpha$ are real $32$-component Majorana spinors that are irreducible under the Lorentz algebra $\mf{so}(1,10)$ and form the representation $R_1=[0,0,0,0,1]$ of the complexified Lie algebra $B_5$. The last label refers to the spinor node of the $B_5$ Dynkin diagram. Carrying out the algorithm~\eqref{eq:recF0} leads to table~\ref{tab:SM11}.

\begin{table}[t!]
\centering
\renewcommand{\arraystretch}{1.3}
\begin{tabular}{c|c|c}
$R_\ell$ & representation & generator\\
\hline\hline
$R_1$ & $[0,0,0,0,1]$ & $Q_\alpha$\\\hline
\multirow{3}{10mm}{\centering$R_2$} & $[1,0,0,0,0]$ & $P_a$\\
& $[0,1,0,0,0]$ & $P_{ab}=P_{[ab]}$\\
& $[0,0,0,0,2]$ & $P_{a_1\ldots a_5}=P_{[a_1\ldots a_5]}$\\\hline
\multirow{5}{10mm}{\centering$R_3$} & $2\times [0,0,0,0,1]$ & spinors\\
& $2\times [1,0,0,0,1]$ & vector-spinors\\
& $[0,1,0,0,1]$ & two-form spinor\\
& $[0,0,1,0,1]$ & three-form spinor\\
& $[0,0,0,1,1]$ & four-form spinor\\\hline
\multirow{1}{10mm}{\centering$R_4$} & $6\times [0,0,0,0,2] $&  $\ldots$\\
& $4\times [0,0,0,1,0]$ & \\
& $ [0,0,0,1,2]$ & \\
& $4\times [0,0,1,0,0]$ & \\
& $2\times [0,0,1,0,2]$ & \\
& $[0,0,1,1,0]$ & \\
& $5\times [0,1,0,0,0]$ & \\
& $2\times [0,1,0,0,2]$ & \\
& $3\times [0,1,0,1,0]$ & \\
& $2\times [0,1,1,0,0]$ & \\
& $3\times [1,0,0,0,0]$ & \\
& $3\times [1,0,0,0,2]$ & \\
& $4\times [1,0,0,1,0]$ & \\
& $4\times [1,0,1,0,0]$ & \\
& $2\times [1,1,0,0,0]$ & \\\hline
$\vdots$& $\vdots$ & $\vdots$
\end{tabular}
\caption{\label{tab:SM11}\it The free Lie superalgebra $\mf{f}(Q)$ in $D=11$ space-time dimensions for a single $32$-component spinor generator $Q_\alpha$. All $\mf{so}(1,10)$ representations are labelled by their $B_5$ Dynkin labels in Bourbaki numbering. Tensor-spinors are gamma traceless but we retain tensor traces. Some representations occur with multiplicity as shown.}
\end{table}

The commutation relations leading to level $\ell=2$ in the free Lie superalgebra $\mf{f}(Q)$ are
\begin{align}
\label{eq:D11}
\lc Q_\alpha , Q_\beta\rc = (C\Gamma^a)_{\alpha\beta}  P_a +  \frac12 (C\Gamma^{ab})_{\alpha\beta}  P_{ab} +  \frac1{5!} (C\Gamma^{a_1\cdots a_5})_{\alpha\beta}  P_{a_1\ldots a_5}\,.
\end{align}
The right-hand side contains the most general combination that can be written for two supersymmetry transformations in $D=11$.  This extension of the Poincar\'e superalgebra was first
considered in\cite{vanHolten:1982mx}, and is nowadays called the M-algebra
(but should not be confused with the generalisation thereof constructed in \cite{Sezgin:1996cj}).
In~\cite{deAzcarraga:1989mza,Townsend:1995gp,Townsend:1997wg} it has been argued that the (non-central under Lorentz) terms $P_{ab}$ and $P_{a_1\cdots a_5}$ beyond the usual translation generator $P_a$ are the ones that correspond to the M2-brane and M5-brane of eleven-dimensional supergravity. The even more general algebra constructed in~\cite{Sezgin:1996cj} contained more brane sources and even fermionic ones that were obtained by treating the $P_a$, $P_{ab}$ and $P_{a_1\ldots a_5}$ as forms in superspace. The immediate interpretation of all these charges was not clear but some were used for studying the Wess--Zumino term of the M5-brane~\cite{Sezgin:1996cj}.

One possible quotient of the free Lie superalgebra $\mf{f}(Q)$ is the standard Poincar\'e superalgebra in $D=11$ (without $P_{ab}$ and $P_{a_1\ldots a_5}$) and another one is the truncation to $\ell=2$ which then reproduces the M-algebra of~\cite{Townsend:1995gp}. Including also generators from higher levels one may construct quotients that agree with the algebra considered in~\cite{Sezgin:1996cj}.

We also note that very recently a supersymmetric extension of the $D=11$ Maxwell algebra was proposed in~\cite{Penafiel:2017wfr,Ravera:2018vra}. This algebra is a quotient of the free Lie superalgebra by retaining the following generators besides the Poincar\'e superalgebra: On $\ell=2$ additionally $P_{ab}$ but not $P_{a_1\ldots a_5}$; on $\ell=3$ only a single spinor $\Sigma_\alpha$ (similar to~\eqref{PQ=Sigma}) and on level $\ell=4$ only a single two-form $Z_{ab}$. Other extensions of the $D=11$ Poincar\'e superalgebra were studied in the context of free differential algebras in~\cite{Andrianopoli:2016osu,Andrianopoli:2017itj}.

\subsubsection{Relation to $E_{11}$ and $\ell_1$ representation?}

It is tempting to speculate on a relation between the free Lie superalgebra and exotic supersymmetric objects as they appear in supergravity and M-theory and are captured by approaches based on the Kac--Moody algebra $E_{11}$~\cite{West:2001as,West:2003fc}. Exotic branes in the sense of~\cite{deBoer:2010ud,deBoer:2012ma} are extended objects of co-dimension at most two and they tend to be very non-perturbative when analysed in string theory, meaning their mass scales with the inverse string coupling to a power that is larger than $1$. These exotic branes have been discovered originally using U-duality in low dimensions (see for instance~\cite{Obers:1998fb,deWit:2000wu}). From the point of view of an electric space-time coupling they typically couple to \textit{mixed symmetry} space-time potentials, meaning potentials that are not necessarily $p$-forms but represented by Young tableaux of more complicated type than a single column. Using this language all such supersymmetric branes have been classified and their relation to non-geometric backgrounds has been studied~\cite{Bergshoeff:2011zk,Kleinschmidt:2011vu,Bergshoeff:2013sxa,Bergshoeff:2015cba}.

Even though exotic branes can typically be expressed as supersymmetric solutions of usual supergravity, meaning that they preserve some of the usual supersymmetries generated by the $Q_\alpha$ at $\ell=1$ in $\mf{f}(Q)$, their (electric) coupling to mixed symmetry potentials suggests to also investigate whether there can be any relation to higher levels $\ell>1$ in the free Lie superalgebra. The motivation from this investigation comes from the fact that the adjoint representation of $E_{11}$ captures all the relevant mixed symmetry potentials~\cite{Kleinschmidt:2011vu} and that the $\ell_1$ representation of $E_{11}$ seems to capture all the corresponding `electric' charges~\cite{West:2003fc,Kleinschmidt:2003jf}. With electric charges we mean that while the M$2$-brane of $D=11$ supergravity couples naturally to the $3$-form potential, the M$5$-brane couples more naturally `electrically' to its dual $6$-form rather than `magnetically' to the original $3$-form. The electric coupling is always such that the corresponding Wess--Zumino term in the world-volume action is simply an integral over (the pull-back of) the corresponding potential. Roughly, for a supersymmetric $p$-brane the closed $p$-form $dX^{a_1} \wedge \cdots\wedge dX^{a_p}$ on the world-volume $\Sigma$ contributes to the supersymmetry algebra if the brane, embedded via the maps $X^a(\xi)$, wraps a topologically non-trivial cycle. This is due to the quasi-invariance of the Wess--Zumino coupling $\int_\Sigma C_{(p+1)}$ to a $(p+1)$-form gauge field $C_{(p+1)}$ under supersymmetry~\cite{deAzcarraga:1989mza}. The contribution to the supersymmetry algebra is of the form
\begin{align}
\lc Q_\alpha, Q_\beta \rc = \frac{1}{p!} (C\Gamma_{a_1\ldots a_p})_{\alpha\beta} Z^{a_1\ldots a_p}\,,\quad\textrm{where }\, Z^{a_1\ldots a_p} \sim Q_p \int d^p\xi\,  dX^{a_1}\wedge \cdots \wedge dX^{a_p}\,,
\end{align}
with the integral over the non-trivial (spatial) cycle that the brane wraps and $Q_p$ the charge of the brane. With this logic the $2$-form $P_{ab}$ in~\eqref{eq:D11} is related to the M$2$-brane while the $5$-form $P_{a_1\ldots a_5}$ is related to the M$5$-brane~\cite{Townsend:1995gp,Townsend:1997wg}. This can be also understood through the gauge-field that the brane couples to as follows. The brane requires a $(p+2)$-form flux in space-time $F_{(p+2)} = d C_{(p+1)}+\ldots$ and in $D$ space-time dimensions the transverse space has a sphere $S^{D-p-2}$ at infinity over which the dual of the flux $F_{(p+2)}$ can be integrated to give the charge $Q_p$ of the extended object. From this we see that the supersymmetry algebra for $p$-brane coupling to a form $C_{a_1\ldots a_{p+1}}$ should contain a (non-central) term $Z_{a_1\ldots a_p}$, i.e., a form with one index less.

As exotic branes couple to mixed symmetry potentials one might therefore by extension wonder whether there are mixed-symmetry `charges' sitting somewhere in an extended supersymmetry algebra to which they couple. As there is no room in the standard $D=11$ superalgebra~\eqref{eq:D11} we look for them in the free Lie superalgebra $\mf{f}(Q)$. In order to identify them we use the following tentative connection to $E_{11}$.

{}From the point of view of $E_{11}$, the generators $P_a$, $P_{ab}$ and $P_{a_1\ldots a_5}$ appear naturally in the so-called $\ell_1$ representation when it is decomposed under $\mf{gl}(11)\subset E_{11}$~\cite{West:2003fc,Kleinschmidt:2003jf}. The relation between space-time potentials in the adjoint of $E_{11}$ and the $\ell_1$ representation has also been discussed in these references. As $E_{11}$ is conjectured to contain all the mixed symmetry space-time potentials that (supersymmetric) branes can couple to, it is an important observation that the $\ell_{1}$ representation of $E_{11}$ contains all the mixed symmetry `charges' that can be obtained by removing one index from any of the $E_{11}$ gauge potentials~\cite{West:2003fc,Kleinschmidt:2003jf,Henneaux:2011mm}.

In order to compare the $\mf{so}(1,10)$ representations predicted by $E_{11}$ and the content of the free Lie superalgebra $\mf{f}(Q)$, we reproduce the lowest levels of the $\ell_1$ representation of $E_{11}$ in terms of $\mf{gl}(11)$ representations in table~\ref{tab:e11l1}. The table also lists their decomposition into $\mf{so}(1,10)$ representations. This should be compared to the \textit{even} levels of the free Lie superalgebras listed in table~\ref{tab:SM11}.

\begin{table}[t!]
\centering
\renewcommand{\arraystretch}{1.3}
\begin{tabular}{c|c|c|c}
Level & $\mf{gl}(11)$ representation & $\mf{so}(1,10)$ representation& Occurrence in $\mf{f}(Q)$\\\hline\hline
$\tfrac32$ & $[1,0,0,0,0,0,0,0,0,0]$ & $[1,0,0,0,0]$& $R_2$\\\hline
$\tfrac52$ & $[0,0,0,0,0,0,0,0,1,0]$ & $[0,1,0,0,0]$& $R_2$ \\\hline
$\tfrac72$ & $[0,0,0,0,0,1,0,0,0,0]$ & $[0,0,0,0,2]$& $R_2$\\\hline
\multirow{3}{10mm}{\centering$\tfrac92$} & \multirow{2}{45mm}{\centering$[0,0,0,1,0,0,0,0,0,1]$} & $[1,0,0,1,0]$& $R_4$\\
&&$[0,0,0,0,2]$& $R_4$\\\cline{2-4}
&$[0,0,1,0,0,0,0,0,0,0]$ & $[0,0,1,0,0]$ & $R_4$\\\hline
\multirow{13}{10mm}{\centering$\tfrac{11}2$} & \multirow{4}{45mm}{\centering$[0,0,1,0,0,0,0,1,0,0]$} & $[0,0,2,0,0]$ & ?\\
&&$[0,1,0,1,0]$ & $R_4$\\
&&$[1,0,0,0,2]$ & $R_4$\\
&&$[0,0,0,0,2]$ & $R_4$\\\cline{2-4}
& \multirow{3}{45mm}{\centering$[0,1,0,0,0,0,0,0,0,2]$} & $[0,1,0,0,0]$ & $R_4$\\
&&$[1,0,1,0,0]$ & $R_4$\\
&&$[2,1,0,0,0]$ & ?\\\cline{2-4}
& \multirow{3}{45mm}{\centering$[0,1,0,0,0,0,0,0,1,0]$} & $[0,0,0,1,0]$ & $R_4$\\
&&$[0,2,0,0,0]$ & $R_4$\\
&&$[1,0,1,0,0]$ & $R_4$\\\cline{2-4}
& \multirow{2}{45mm}{\centering$2\times [1,0,0,0,0,0,0,0,0,1]$} & $2\times [0,1,0,0,0]$ & $R_4$\\
&&$2\times [2,0,0,0,0]$ & ?\\\cline{2-4}
& $[0,0,0,0,0,0,0,0,0,0]$ & $[0,0,0,0,0]$ &?\\\hline
$\vdots$& $\vdots$ & $\vdots$ & $\vdots$
\end{tabular}
\caption{\label{tab:e11l1}\it The first few levels of the $\ell_1$ lowest weight representation of $E_{11}$. Level here refers to the natural grading under $\mf{gl}(11)\subset E_{11}$. The last column lists a possible match with the even levels of the free Lie algebra in table~\ref{tab:SM11}.}
\end{table}

The comparison has been carried out at the level of representations up to $R_4$ in the last column of table~\ref{tab:e11l1}. As can be seen from that column, most of the mixed symmetry representations contained in the $\ell_1$ representation arise already on the first two bosonic levels $R_2$ and $R_4$ of the free Lie superalgebra; it is possible that the question marks will be contained in the higher levels. As the charges in the $\ell_1$ representation cover \textit{all} branes, the analysis above includes standard and exotic branes. 

We wish to stress that this comparison is speculative and we have only analysed the representations. The algebraic structure is quite different, however. In the $E_{11}$ proposal of~\cite{West:2003fc}, all generators in $\ell_1$ commute: they form a generalised abelian translation algebra of a generalised space-time in a way similar to the generalised diffeomorphisms of exceptional geometry~\cite{Berman:2011cg,Berman:2011jh,Coimbra:2011ky,Berman:2012vc}. One of the hallmarks of the Maxwell algebra and free Lie algebra approach pursued here is that the translations $P_a$ no longer commute, see~\eqref{eq:PPZ}. This can be achieved in an $E_{11}$-covariant fashion by allowing non-trivial commutation relations for the elements of the $\ell_1$ representation. One possibility would be an embedding in $E_{12}$ along the lines of~\cite{Kleinschmidt:2003jf}, see also~\cite{Kleinschmidt:2003pt}. This embedding embeds the semi-direct sum $E_{11}\oplus \ell_1$ as levels $0$ and $1$ of a graded decomposition of $E_{12}$ under its obvious $E_{11}$ subalgebra. We shall make more comments on such level decompositions in the next section. Another possibility is to consider the elements in the $\ell_1$ representation as odd such that they satisfy non-trivial anti-commutation relations rather than commutation relations. This structure can be embedded in
the Borcherds superalgebra denoted $\mathcal{B}_{11}$ in \cite{Kleinschmidt:2013em}, or the tensor hierarchy algebra considered in \cite{Bossard:2017wxl}.
We refrain from speculating further in this direction, but we will in the next section apply the construction of a Borcherds superalgebra to the case
considered in section~\ref{sec:examples}
where the basic objects are the supersymmetry generators $Q_\alpha$. 

\section{Free Lie superalgebras and Borcherds superalgebras}
\label{sec:Borcherds}

We have seen that a free Lie (super)algebra decomposes into a direct sum of subspaces at positive levels, and that we can add
a Lie algebra $\mf{g}_0$ at level zero, such that the positive level subspaces form representations of it.
In order to obtain a simple Lie (super)algebra we need to continue to negative levels
since the subspace spanned by all elements at levels $\ell>k$ for a given $k\geq1$ otherwise constitutes an ideal, as described in the end of
section~\ref{sec:MaxwellFree}.
Any finite-dimensional complex semisimple Lie algebra has this structure, with the Cartan subalgebra at level zero and root vectors corresponding to positive and negative roots at positive and negative levels.
The construction of the algebra from its Cartan matrix or Dynkin diagram can be generalised in different ways, leading to infinite-dimensional Kac--Moody algebras, the Borcherds superalgebras that we will describe next, and most generally, to so-called contragredient Lie (super)algebras \cite{Kac68,Kac77B}.
As we will show, when $\mf{g}_0$ is semisimple one can extend its Cartan matrix or Dynkin diagram and apply the generalised construction in order to obtain a corresponding extension of $\mf{g}_0$ with
two subalgebras at positive and negative levels, respectively, which are quotients of two (isomorphic) free Lie (super)algebras,
and additional Cartan generators at level zero.

A Borcherds superalgebra of rank $r$ is defined from a symmetric
Cartan matrix $A_{ij}$ ($i,j=1,2,\ldots,r$) with non-positive off-diagonal entries and a subset $S$ of the index set
$\{1,2,\ldots,r\}$
labelling the rows and columns. Here we will only consider the case when the Cartan matrix is non-degenerate, integer-valued and satisfies
\begin{align}
\label{eq:CMcond}
i \notin S &\Leftrightarrow A_{ii}=2\,, & i \in S &\Leftrightarrow A_{ii}=0\,.
\end{align}
To the Cartan matrix we associate $3r$ generators $e_i, h_i, f_i$ among which $e_i,f_i$ are odd if and only if $i \in S$, and all $h_i$ are even.
The Borcherds superalgebra is now defined as the Lie superalgebra generated by $e_i, h_i, f_i$ modulo the relations
\begin{subequations}\label{eq:chevserre}
\begin{align}
\label{eq:BA1}
\lb h_i, e_j \rb &= A_{ij} e_j\,,\quad \lb h_i, f_j \rb = -A_{ij} f_j\,,\quad \lbb e_i, f_j\rbb = \delta_{ij} h_i\,, \quad \lb h_i, h_j\rb =0\,,\\
\label{eq:serre}
&\quad i\neq j\quad\!\!\textrm{and $i\notin S$} \quad \Rightarrow \quad(\ad e_i)^{1-A_{ij}} e_j = 0\,,\quad (\ad f_i)^{1-A_{ij}} f_j = 0\,,\\
\label{eq:ferm2}
&\quad\quad\qquad A_{ii}=0 \quad \Rightarrow \quad \{ e_i , e_i\} =0\,,\quad \{ f_i , f_i\} =0\,.
\end{align}
\end{subequations}
We have used the notation $\lbb\cdot,\cdot\rbb$ to denote the graded commutator since the $e_i$ and $f_i$ can be odd or even.
The off-diagonal entries of the Cartan matrix $A_{ij}$ determine the Serre relations of~\eqref{eq:serre}. The adjoint action there is also by the graded commutator.

The Dynkin diagram that we associate to the Cartan matrix consists of $r$ nodes, where node $i$ ($i=1,2,\ldots,r$) is white (depicted with $\bigcirc$) if 
$i \notin S \Leftrightarrow A_{ii}=2$ and gray (depicted with $\otimes$) if $i \in S \Leftrightarrow A_{ii}=0$,
and where two nodes $i$ and $j$ are connected with $-A_{ij}$ lines. (We note that the relation between Cartan matrices and Dynkin diagrams is not one-to-one for superalgebras as their are odd Weyl reflections that can be applied~\cite{Dobrev}.)

As noted above, Borcherds superalgebras have a triangular decomposition: They consist of the Cartan subalgebra spanned by the $h_i$, a lower-triangular part generated by the $f_i$ through multi-commutation and an upper-triangular part generated by the $e_i$ through multi-commutation. Important for us is the observation that the upper-triangular part is a free Lie superalgebra modulo the Serre relations~\eqref{eq:serre} and~\eqref{eq:ferm2}. In other words, the Serre relations define an ideal in a free Lie superalgebra that has to be quotiented out. 
We can refine the observation by allowing each generator $e_i$ to have an arbitrary non-negative level $v_i$, and the corresponding $f_i$ to have level $-v_i$. If $v_i=1$ for all $i$, then we get the grading described above, with only the Cartan subalgebra at level zero. But we can also choose to have $v_i=1$ only for some $i$, and $v_i=0$ for the others. 
By setting $v_i=1$ if $i \in S$ and $v_i=0$ if $i \notin S$ we get a $\mathbb{Z}$-grading consistent with the $\mathbb{Z}_2$-grading: odd elements appear at odd levels, and even elements at even levels. The level $\ell=0$ subalgebra $\mf{g}_0$ is then a semisimple Kac--Moody algebra with a Dynkin diagram given by removing the gray nodes from the Dynkin diagram of the Borcherds superalgebra, centrally extended with the Cartan generators $h_i$ of the removed nodes. In this level decomposition, the Serre relations \eqref{eq:serre} determine the level $\ell=0$ subalgebra $\mf{g}_0$ and its representations at levels $\ell=\pm1$ (which will be dual to each other), whereas the additional Serre relations \eqref{eq:ferm2}
define the ideal that is factored out from the free Lie superalgebras at positive and negative levels
generated by the subspaces at level $\ell=1$ and $\ell=-1$, respectively.

\subsection{$D=4$ and $\cN=1$ supersymmetry} 

\begin{figure}[t!]
\centering
\begin{picture}(130,30)
\thicklines
\put(10,20){\circle{7}}
\put(13.5,20){\line(1,0){30.2}}
\put(47,20){\circle{7}}
\put(47,20){\line(1,1){2.5}}
\put(47,20){\line(-1,1){2.5}}
\put(47,20){\line(1,-1){2.5}}
\put(47,20){\line(-1,-1){2.5}}
\put(50.6,21){\line(1,0){30.2}}
\put(50.6,19){\line(1,0){30.2}}
\put(84,20){\circle{7}}
\put(84,20){\line(1,1){2.5}}
\put(84,20){\line(-1,1){2.5}}
\put(84,20){\line(1,-1){2.5}}
\put(84,20){\line(-1,-1){2.5}}
\put(87.7,20){\line(1,0){30.2}}
\put(121.7,20){\circle{7}}
\put(7.5,5){$1$}
\put(44,5){$2$}
\put(82,5){$3$}
\put(119,5){$4$}
\end{picture}
\caption{\label{fig:SM4}\it Dynkin diagram of a Borcherds Lie superalgebra that is related to a quotient of the free Lie superalgebra of $D=4$ supersymmetry.}
\end{figure}
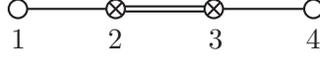

We will now try to reconstruct an infinite-dimensional extension of the superalgebra considered in~\eqref{eq:BGKL} as the subalgebra at non-negative levels of a Borcherds superalgebra.
In agreement with the discussion above we should have at level $\ell=0$ the complexification $A_1\oplus A_1$ of the
Lorentz algebra $\mf{so}(1,3)$ in $D=4$ and at level $\ell=1$ the spinors $[0,1]\oplus[1,0]$. One way to arrange this is by taking the Cartan matrix
\begin{align}
\label{eq:SM4A}
A= \begin{pmatrix} 
2 & -1 & 0 & 0 \\
-1 & 0 & -2 & 0 \\
0 & -2 & 0 &-1 \\
0 & 0 & -1 & 2
\end{pmatrix}\,.
\end{align}
This can also be depicted in terms of a Dynkin diagram as in figure~\ref{fig:SM4}.

The two $A_1$ forming the Lorentz algebra in $D=4$ correspond to nodes $1$ and $4$ of the Dynkin diagram and the gray nodes are fermionic and represent the two fundamental representations of the two $A_1$ as required by the fundamental spinor $Q_\alpha \leftrightarrow [1,0]\oplus[0,1]$.
Thus the subset $S$ of the index set $\{1,2,3,4\}$ is $S=\{2,3\}$ here.
The single lines connecting the gray nodes to nodes $1$ and $4$ imply that the corresponding Chevalley generators belong to a two-dimensional fundamental representation of the
$A_1$ concerned.

Let us first see how to get the Lorentz algebra by removing the gray nodes. If we set
\begin{subequations}\label{eq:Mef}
\begin{align}
M_{02}&=\frac{i}{2}(-e_1+f_1+e_4-f_4)\,,&  M_{01}&=\frac12(-e_1-f_1-e_4-f_4)\,,\\ 
M_{23}&=\frac{i}{2}(e_1+f_1-e_4-f_4)\,,&  M_{31}&=\frac12(-e_1+f_1-e_4+f_4)\,,\\ 
M_{21}&=\frac{i}{2}(h_1-h_4)\,, &  M_{03}&=\frac12(h_1+h_4)\,,
\end{align}
\end{subequations}
the commutation relations (\ref{eq:MMM}) indeed follow from (\ref{eq:chevserre}) in the case where the Cartan matrix $A_{ij}$ is given by (\ref{eq:SM4A}), or equivalently,
by the Dynkin diagram in figure \ref{fig:SM4}.
Note that the imaginary unit $i$ appears in (\ref{eq:Mef}) although we consider the real Lie algebra $\mf{so}(1,3)$
(no $i$ appears in the commutation relations (\ref{eq:MMM})).
This shows that $\mf{so}(1,3)$ is a non-split real from of $A_1 \oplus A_1$. If we instead just took the real span of $e_1,f_1,h_1,e_4,f_4,h_4$, then we would get
the split real form $\mf{sl}(2)\oplus\mf{sl}(2)$ of $A_1 \oplus A_1$. Note also that $e_2,f_2,h_2,e_3,f_3,h_3$ do not appear in (\ref{eq:Mef})
since we have removed the gray nodes. However, the Cartan generators
$h_2$ and $h_3$ do belong to level zero as well and by taking appropriate linear combinations of them,
one can get the dilatation operator $d$ and the chirality operator $B_5$ that also appears in the algebra~\eqref{eq:BGKL}.

We now put back the gray nodes and go from level $\ell=0$ to $\ell=1$.
Looking for example at the $e_2$ Chevalley generator, the Serre relation implies
\begin{align}
e_2 \neq 0\,,\quad
\lb e_1 , e_2 \rb \neq 0\,,\quad\textrm{but} \quad
\lb e_1, \lb e_1, e_2\rb \rb =0\,,
\end{align}
so that there are two non-zero elements at level $\ell=1$ associated with $e_2$. This is the $[1,0]$ representation of $A_1\oplus A_1$.
One can verify the Dynkin labels $[1,0]$ by acting with the Cartan generators $h_1$ and $h_4$ on the highest weight vector $[e_1,e_2]$:
\begin{align}
[h_1,[e_1,e_2]]&=1\,,&
[h_4,[e_1,e_2]]&=0\,.
\end{align}
A similar reasoning for nodes $3$ and $4$ yields the $[0,1]$ representation. The level $\ell$ here is given by the sum of the number of times $e_2$ or $e_3$ appear in a multi-commutator.
Thus one of the two 2-dimensional Weyl spinors at level $\ell=1$ is spanned (over the complex numbers) by $e_2$ and $[e_1,e_2]$ and the other one by
$e_4$ and $[e_3,e_4]$. In order to get a Majorana spinor $Q_\alpha$ ($\alpha=1,2,3,4$) from them, we have to take the complex linear combinations
\begin{subequations}
\begin{align}
Q_1 &= [e_1,e_2]-i[e_3,e_4]\,,\\ 
Q_2 &= i[e_1,e_2]-[e_3,e_4]\,,\\ 
Q_3 &= e_2+ie_3\,,\\ 
Q_4 &= -ie_2 - e_3\,.
\end{align}
\end{subequations}
Indeed, now the commutation relations (\ref{eq:MQ}), which state that $Q_\alpha$ transforms as a Majorana spinor, follow
from (\ref{eq:chevserre}), and the explicit form of the real gamma matrices given in \ref{gammapp}.

We proceed to level $\ell=2$ and set
\begin{subequations}
\begin{align}
P_0 &= i [e_1,\{e_2,[e_3,e_4]\}]-i\{e_2,e_3\}\,,\\ 
P_1 &= i [e_1,\{e_2,e_3\}]-i\{e_2,[e_3,e_4]\}\,,\\ 
P_2 &= - [e_1,\{e_2,e_3\}]-\{e_2,[e_3,e_4]\}\,,\\ 
P_3 &= i [e_1,\{e_2,[e_3,e_4]\}]+i\{e_2,e_3\}\,.
\end{align}
\end{subequations}
Note that in each multi-commutator, $e_2$ and $e_3$ appear once each. A  multi-commutator with two $e_2$ and no $e_3$ (or two $e_3$ and no $e_2$) would also belong to level $\ell=2$, but any such element is zero because of the Serre relations
\begin{align} \label{eq:Serre2002}
\{e_2,e_2\}=\{e_3,e_3\}=0\,.
\end{align}
This means that $P_{ab}$, that appeared in the anti-commutation relation (\ref{eq:QQfree}) of the free Lie superalgebra
is not present in the Borcherds superalgebra we consider here,
since $\{e_2,e_2\}$ and $\{e_3,e_3\}$ are lowest weight vectors of the representation $[2,0]$ and $[0,2]$, respectively,
which together form $P_{ab}$. Rather we get (\ref{eq:QQ}) 
from (\ref{eq:chevserre}).
Also the relation (\ref{eq:MP}), which states that $P_a$ transforms as vector under the Lorentz algebra, follow from (\ref{eq:chevserre}).\footnote{\label{fn:black}At this point, we note that there is the possibility of writing down a Borcherds algebra whose positive levels agrees with the free Lie superalgebra including $P_{ab}$. In order to arrange for this, one has to consider more general Cartan matrices than the ones we have introduced in~\eqref{eq:CMcond} by also allowing negative diagonal entries, e.g. $A_{ii}=-2$. The defining relations can for instance be found in~\cite{Ray} and the corresponding nodes are sometimes called `black' nodes. For black nodes, the relation~\eqref{eq:ferm2} does not apply and therefore there are no Serre relations at positive levels associated with the black nodes, in other words they form a free Lie superalgebra.}

In drawing diagram~\ref{fig:SM4} we have chosen a double line between nodes $2$ and $3$,
corresponding to $A_{23}=A_{32}=-2$. For understanding the Serre relations 
drawing a single or multiple lines between the nodes does not matter, as any negative choice of $A_{23}$ implies the same ideal. This is due to the Jacobi identity
\begin{align}
\lb e_2, \lc e_2, e_3\rc\rb = \lb \lc e_2, e_2\rc, e_3\rb -\lb e_2, \lc e_2, e_3\rc\rb \,.
\end{align}
The first term on the right-hand side vanishes due to~\eqref{eq:ferm2} and the sign in front of the second term follows from the oddness of $e_2$. This 
Jacobi identity therefore implies
\begin{align}
\label{eq:serre4}
\lb e_2, \lc e_2, e_3\rc\rb  =0 
\end{align}
independent of the value of $A_{23}$. The value of $A_{23}$ does, however, lead to inequivalent Borcherds superalgebras when one considers the action of the lower-triangular Chevalley generators $f_i$ as well. As long as one is only interested in the upper triangular part, $A_{23}$ does not matter. As we have just shown,~\eqref{eq:serre4} is a consequence of $\lc e_2, e_2 \rc =0$. In the free Lie superalgebra, this relation is not imposed. 
We thus conclude that the upper-triangular part of the Borcherds superalgebra defined by~\eqref{eq:SM4A} is a free Lie superalgebra on the $A_1\oplus A_1$ representations $[1,0]\oplus[0,1]$ modulo the Serre relations~\eqref{eq:Serre2002} corresponding to an ideal
generated by $[2,0] \oplus [0,2]$ at level $\ell=2$.

The representation content of the positive levels of the Borcherds superalgebra can be constructed recursively using a similar algorithm as for the free Lie superalgebra. In the Borcherds superalgebra one has
\begin{subequations}
\label{eq:recF}
\begin{align}
B_1 &= \left\langle Q_\alpha \right\rangle\,,\\
B_2 &= \SYM^2 B_1\ominus S_2\,,\\
B_3 &= B_2 \otimes B_1 \ominus \SYM^3 B_1\oplus S_3 \,,\\
B_4 &= (B_3 \otimes B_1 \oplus \ALT^2 B_2) \ominus \SYM^2 B_1\otimes B_2 \oplus \SYM^4 B_1 \ominus S_4\,,\\
&\cdots \nn
\end{align}
\end{subequations}
where the difference to the free Lie superalgebra~\eqref{eq:recF0} is the presence of a certain representation
$S_\ell$ at each level that has to be removed. We have already seen that $S_2=[2,0]\oplus[0,2]$ and an analysis 
similar to the one carried out in \cite{Cederwall:2015oua} for the case of only one gray node
yields that in general $S_\ell=[\ell,0]\oplus[0,\ell]$. This is displayed in table~\ref{tab:SM4B}.

\begin{table}[t!]
\centering
\renewcommand{\arraystretch}{1.3}
\begin{tabular}{c|c|c}
$B_\ell$ & representation & generator\\
\hline\hline
\multirow{3}{10mm}{\centering$B_0$} & $[2,0]\oplus [0,2]$ & $M_{ab}$\\
& $[0,0]$ & $B_5$ \\
& $[0,0]$ & $d$\\\hline
$B_1$ & $[1,0]\oplus [0,1]$ & $Q_\alpha$\\\hline
$B_2$ & $[1,1]$ & $P_a$\\\hline
$B_3$ & $[1,0]\oplus [0,1]$ & $\Sigma_\alpha$\\\hline
\multirow{2}{10mm}{\centering$B_4$} & $[2,0]\oplus [0,2]$ & $Z_{ab}=Z_{[ab]}$\\
& $[0,0]$ & $B$\\
\hline
\multirow{2}{10mm}{\centering$B_5$} & $[1,0]\oplus [0,1]$ & $X_\alpha$\\
& $[1,2]\oplus [2,1]$ & $X_{a\alpha}$\\
\hline
\multirow{5}{10mm}{\centering$B_6$} & $[2,0]\oplus [0,2]$ & \\
 & $[1,1]$& \\  & $[1,1]$& \\  & $[1,1]$& \\  & $[1,3]\oplus [3,1]$& \\ \hline
 $\vdots$& $\vdots$ & $\vdots$
\end{tabular}
\caption{\label{tab:SM4B}\it The non-negative levels of the Borcherds superalgebra defined by the Cartan matrix~\eqref{eq:SM4A}. This is a quotient of the free Lie superalgebra shown in table~\ref{tab:SM4} and the notation is in that table. We have also included the Cartan generators from level $\ell=0$. $B_5$ acts as chirality while $d$ is a scaling operator. Up to level $\ell=4$ this is the Maxwell superalgebra studied in \cite{Bonanos:2009wy}.}
\end{table}

Alternatively, the representation content of the Borcherds superalgebra up to an arbitrary level $k$ 
can be derived from the representation content of a corresponding Kac--Moody algebra. This can for example be more useful than the method above when
the representations $S_\ell$ are not known.
This `Borcherds-Kac-Moody correspondence' was first explained in \cite{Henneaux:2010ys} in the context of $E_{11}$ (see also \cite{Palmkvist:2011vz})
and then generalised to other cases in \cite{Palmkvist:2012nc,Howe:2015hpa}. Here we will just briefly describe the
consequence of the general procedure in the special case of the Borcherds superalgebra with Cartan matrix \eqref{eq:SM4A}
or with the Dynkin diagram in figure \ref{fig:SM4}, and it reveals a remarkable connection between the Maxwell superalgebra and the 
exceptional Lie algebra $E_6$. The method is only directly applicable if $A_{23}=A_{32}=-2$, and this is the reason why we have made this choice, although, as we have seen, any negative value of these entries in the Cartan matrix gives the same representation contents in the
level decomposition.

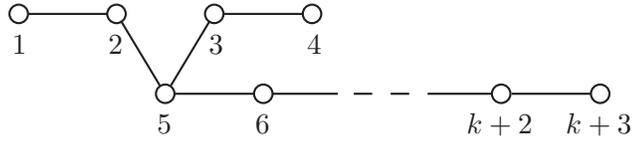
\begin{figure}[t!]
\centering
\begin{picture}(227,70)(6.5,-33.7)
\thicklines
\put(10,20){\circle{7}}
\put(13.5,20){\line(1,0){30.2}}
\put(47,20){\circle{7}}
\put(48.7,17.4){\line(4,-6.65){14.7}}
\put(82.3,17.4){\line(-4,-6.65){14.7}}
\put(84,20){\circle{7}}
\put(65.5,-10.2){\circle{7}}
\put(69,-10.2){\line(1,0){30.2}}
\put(102.5,-10.2){\circle{7}}
\put(106,-10.2){\line(1,0){24.7}}
\put(136.7,-10.2){\line(1,0){7}}
\put(150.7,-10.2){\line(1,0){7}}
\put(164.7,-10.2){\line(1,0){24.7}}
\put(192.5,-10.2){\circle{7}}
\put(196.5,-10.2){\line(1,0){30.2}}
\put(230,-10.2){\circle{7}}
\put(87.7,20){\line(1,0){30.2}}
\put(121,20){\circle{7}}
\put(7.5,5){$1$}
\put(44,5){$2$}
\put(82,5){$3$}
\put(119,5){$4$}
\put(62.5,-25.2){$5$}
\put(99.5,-25.2){$6$}
\put(179.5,-25.2){$k+2$}
\put(217,-25.2){$k+3$}
\end{picture}
\caption{\it Dynkin diagram of the Kac--Moody algebra of rank $k+3$ corresponding to the Borcherds superalgebra with the Dynkin diagram in
figure \ref{fig:SM4} by the `Borcherds--Kac--Moody correspondence'. The number of nodes in the lower line is
$k-1$, forming an $A_{k-1}$ subdiagram. For $k=3$, this Kac--Moody algebra is the exceptional Lie algebra $E_6$, for $k=4$ its affine extension $E_6{}^+$,
and for $k=5$ its hyperbolic extension $E_6{}^{++}$.\label{fig:extendeddynkin}}
\end{figure}

\begin{table}[t!]
\centering
\renewcommand{\arraystretch}{1.3}
\begin{tabular}{c|c|c}
level $\ell$& $A_1 \oplus A_1 \oplus A_4$ representation & generator\\
\hline\hline
\multirow{4}{10mm}{\centering$0$} & $[2,0;0,0,0,0]\oplus [0,2;0,0,0,0]$ & $M_{ab}$\\
& $[0,0;0,0,0,0]$ & $B_5$ \\
& $[0,0;0,0,0,0]$ & $d$\\
& $[0,0;1,0,0,1] $&\\
\hline
$1$ & $[1,0;1,0,0,0]\oplus [0,1;1,0,0,0]$ & $Q_\alpha{}^{I}$\\\hline
$2$ & $[1,1;0,1,0,0]$ & $P_a{}^{IJ}$\\\hline
$3$ & $[1,0;0,0,1,0]\oplus [0,1;0,0,1,0]$ & $\Sigma_\alpha{}{}^{IJK}$\\\hline
\multirow{3}{10mm}{\centering$4$} & $[2,0;0,0,0,1]\oplus [0,2;0,0,0,1]$ & $Z_{ab}{}^{I_1\ldots I_4}$\\
& $[0,0;0,0,0,1]$ & $B{}^{I_1\ldots I_4}$\\
& $[0,0;1,0,1,0]$ & \\
\hline
\multirow{3}{10mm}{\centering$5$} & $[1,0;0,0,0,0]\oplus [0,1;0,0,0,0]$ & $X_\alpha{}^{I_1\ldots I_5}$\\
& $[1,2;0,0,0,0]\oplus [2,1;0,0,0,0]$ & $X_{a\alpha}{}^{I_1\ldots I_5}$\\
& $[1,0;1,0,0,1]\oplus [0,1;1,0,0,1]$ & \\
\hline
$\vdots$& $\vdots$ & $\vdots$
\end{tabular}
\caption{\label{tab:SM4A}\it The non-negative levels of $E_6{}^{++}$ (the hyperbolic extension of $E_6$)
with Dynkin diagram in figure \ref{fig:extendeddynkin} for $k=5$, decomposed with respect to the $A_1 \oplus A_1 \oplus A_4$ subalgebra obtained by 
removing node 2 and 3. The Dynkin labels of $A_4$ are separated from those of $A_1 \oplus A_1$ with a semicolon.
Restricting to $A_4$ representations where
the $\ell$-th Dynkin label is equal to one and the others to zero (all zero for $\ell=0$ and $\ell=5$) we obtain the $A_1 \oplus A_1$
representations $B_\ell$ in table~\ref{tab:SM4}. We have illustrated this by putting $\ell$ fully antisymmetric indices $I,J,\ldots=1,\ldots,5$ on the generators in table~\ref{tab:SM4}
up to $\ell=5$ (which coincide with those of the Maxwell superalgebra up to $\ell=4$).}
\end{table}

The Dynkin diagram of the corresponding Kac--Moody algebra is obtained 
by replacing the gray nodes with ordinary white ones, removing the double line, and adding another $k-1$ white nodes, each connected to the previous one with
a single line, so that they from the Dynkin diagram of the Lie algebra $A_{k-1}$ with split real form $\mathfrak{sl}(k)$.
Furthermore, the new nodes 2 and 3 replacing the gray ones should be 
connected to the first node in this $A_{k-1}$ subdiagram. The resulting Dynkin algebra is shown in figure \ref{fig:extendeddynkin}. For $k=3$,
the corresponding Kac--Moody algebra is the exceptional Lie algebra $E_6$, for $k=4$ its affine extension $E_6{}^+$,
and for $k=5$ its hyperbolic extension $E_6{}^{++}$.
A level decomposition of this Kac--Moody algebra
can be performed with respect to nodes 2 and 3 in the same way as for the original Borcherds superalgebra,
and since we now have a Lie algebra rather than a Lie superalgebra, the representations can be computed easily using the software SimpLie \cite{SimpLie}.
Furthermore, since we now at level $\ell=0$ have not only $A_1 \oplus A_1$ and the Cartan generators $h_2$ and $h_3$, but also $\mf{sl}(k)$,
each representation of $A_1 \oplus A_1$ comes together with a representation of $\mf{sl}(k)$. 
It turns out that, at each level $\ell\leq k$,
the $A_1 \oplus A_1$ representation that comes together with the $k$-th antisymmetrised tensor power of the fundamental $k$-dimensional representation 
of $\mf{sl}(k)$ is precisely the representation $B_\ell$ appearing in the Borcherds superalgebra. This can be verified for $k=5$ by comparing tables \ref{tab:SM4B} and \ref{tab:SM4A}.
Using either tool we obtain
the supersymmetric Maxwell algebra studied in~\cite{Bonanos:2009wy} with an additional scaling operator $d$ that can be chosen to have the level $\ell$ as its eigenvalue.

\subsection{$D=4$ and extended supersymmetry}

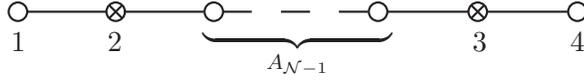
\begin{figure}[t!]
\centering
\begin{picture}(227,50)
\thicklines
\put(10,20){\circle{7}}
\put(13.5,20){\line(1,0){30.2}}
\put(47,20){\circle{7}}
\put(47,20){\line(1,1){2.5}}
\put(47,20){\line(-1,1){2.5}}
\put(47,20){\line(1,-1){2.5}}
\put(47,20){\line(-1,-1){2.5}}
\put(184,20){\circle{7}}
\put(184,20){\line(1,1){2.5}}
\put(184,20){\line(-1,1){2.5}}
\put(184,20){\line(1,-1){2.5}}
\put(184,20){\line(-1,-1){2.5}}
\put(187.7,20){\line(1,0){30.2}}
\put(221.7,20){\circle{7}}
\put(50,20){\line(1,0){30.2}}
\put(180.4,20){\line(-1,0){30.2}}
\put(84.1,20){\circle{7}}
\put(146.3,20){\circle{7}}
\multiput(87.5,20)(22,0){3}{\line(1,0){11}}
\put(7.5,5){$1$}
\put(44,5){$2$}
\put(182,5){$3$}
\put(219,5){$4$}
\put(80,15){$\underbrace{\phantom{sdjflisdjdsfsd}}_{A_{\cN-1}}$}
\end{picture}
\caption{\label{fig:BorcherdsN}\it Dynkin diagram of the Borcherds superalgebra  for $\cN$-extended supersymmetry. The middle part is the Dynkin diagram of type $A_{\cN-1}$ corresponding to the complexified $\mf{su}(\cN)$ R-symmetry. The outer nodes labelled $1$ and $4$ correspond to $A_1\oplus A_1$ as before while the two gray nodes correspond to the two spinors.}
\end{figure}

The Borcherds superalgebras considered in the preceding subsection can be extended in a way corresponding to extended supersymmetry. In order to get an
R-symmetry
algebra $\mathfrak{su}(\mathcal N)$
at level $\ell=0$ (commuting with the Lorentz algebra), and the two Weyl spinors at level $\ell=1$
transforming in two $\mathcal N$-dimensional representations of $\mathfrak{su}(\mathcal N)$, dual to each other (`fundamental' and `anti-fundamental'), we insert 
$\mathcal N-1$ white nodes between the two gray nodes in the Dynkin diagram, as shown in figure \ref{fig:BorcherdsN}. The generators $e_2$ and $e_3$ are then still lowest weight vectors of two representations of the subalgebra at level $\ell=0$, but since this subalgebra now also contains $A_{\mathcal N-1}$ in addition to the (complexified) Lorentz algebra $A_1 \oplus A_1$, the representations now have $1+1+(\mathcal N -1)$ Dynkin labels. As in section~\ref{sec:D4ext} we have put the Dynkin labels corresponding to $A_{\mathcal N-1}$ after the two Dynkin labels corresponding to $A_1 \oplus A_1$, separated by a semicolon. The Dynkin labels of the two representations at level $\ell=1$ are then $[1,0;1,0,\ldots,0]$ and $[0,1;0,\ldots,0,1]$, respectively, as can be verified by acting on $e_2$ and $e_3$ with the Cartan generators.

We also give the Cartan matrix corresponding to the simplest extended case $\cN=2$ for which there are five nodes in figure~\ref{fig:BorcherdsN}; it is
\begin{align}
\label{eq:CMN2}
\begin{pmatrix}
2 & -1 & 0 &0 &0\\
-1 & 0 & 0 &0 & -1\\
0 & 0  & 0 &-1 & -1\\
0 & 0 & -1 &2 &0\\
0 &-1 & -1 & 0 &2
\end{pmatrix}\,.
\end{align}
Row and column $5$ correspond to the R-symmetry $A_1$ in the middle of the diagram.

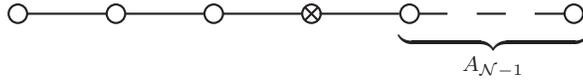
\begin{figure}[t!]
\centering
\begin{picture}(227,50)(-66.5,0)
\thicklines
\put(-64,20){\circle{7}}
\put(-27,20){\circle{7}}
\put(10,20){\circle{7}}
\put(-60.5,20){\line(1,0){30.2}}
\put(-23.5,20){\line(1,0){30.2}}
\put(13.5,20){\line(1,0){30.2}}
\put(47,20){\circle{7}}
\put(47,20){\line(1,1){2.5}}
\put(47,20){\line(-1,1){2.5}}
\put(47,20){\line(1,-1){2.5}}
\put(47,20){\line(-1,-1){2.5}}
\put(50,20){\line(1,0){30.2}}
\put(84.1,20){\circle{7}}
\put(146.3,20){\circle{7}}
\multiput(87.5,20)(22,0){3}{\line(1,0){11}}
\put(80,15){$\underbrace{\phantom{sdjflisdjdsfsd}}_{A_{\cN-1}}$}
\end{picture}
\caption{\label{fig:BorcherdsNdist}\it Distinguished Dynkin diagram of the finite-dimensional contragredient Lie superalgebra $A(3,\cN-1)$, which has the superconformal algebra $\mathfrak{su}(2,2|\cN)$ in $D=4$ as a real form.}
\end{figure}

At this point we note a certain similarity to the superconformal algebra $\mathfrak{su}(2,2|\cN)$ in $D=4$.
This is a real form of the complex Lie superalgebra $A(3,\cN-1)$, which has a distinguished Dynkin diagram shown in figure \ref{fig:BorcherdsNdist}.
The fact that the Dynkin diagram is distinguished means that there is only one gray node, and this condition makes it unique. By relaxing this condition and performing so-called odd reflections \cite{Dobrev} one can obtain equivalent non-distinguished Dynkin diagrams of the same Lie superalgebra, with more than one gray node. Following the rules for odd reflections (see for example \cite{Frappat}) it is straightforward to show that one of the possible diagrams look the same as the one in figure \ref{fig:BorcherdsN}~\cite{Dobrev,Beisert:2003yb}. However, the meaning of gray nodes in this context is slightly different. The corresponding $e$ and $f$ generators are still odd elements, and the corresponding diagonal entries are still zero, but on one side of the gray node the corresponding entries in the Cartan matrix have opposite signs.
For figure \ref{fig:BorcherdsNdist}, this means that if the white nodes and lines on the left hand side of the gray node correspond to diagonal entries $2$ and off-diagonal entries $-1$, respectively,
then those on the right hand side have diagonal entries $-2$ and off-diagonal entries $1$, respectively.
For figure \ref{fig:BorcherdsN}, with $\cN=2$, this gives the corresponding Cartan matrix 
\begin{align}
\label{eq:CMN3}
\begin{pmatrix}
2 & -1 & 0 &0 &0\\
-1 & 0 & 0 &0 & 1\\
0 & 0  & 0 &-1 & 1\\
0 & 0 & -1 &2 &0\\
0 &1 & 1 & 0 &-2
\end{pmatrix}\,,
\end{align}
of $A(3,1)$, where the entries in row or column 5 have opposite sign compared to \eqref{eq:CMN2}. Accordingly, $A(3,1)$ is not a Borcherds superalgebra but a contragredient Lie superalgebra \cite{Kac77B}, and this change of signs makes it finite-dimensional.

\subsection{$D=11$ and $\mathcal N=1$ supersymmetry}

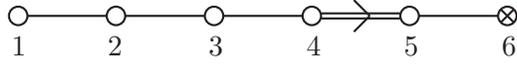
\begin{figure}[t!]
\centering
\begin{picture}(210,30)
\thicklines
\multiput(10,20)(37,0){6}{\circle{7}}
\multiput(13.5,20)(37,0){3}{\line(1,0){30.2}}
\put(195,20){\line(1,1){2.5}}
\put(195,20){\line(-1,1){2.5}}
\put(195,20){\line(1,-1){2.5}}
\put(195,20){\line(-1,-1){2.5}}
\put(124.6,21){\line(1,0){30.2}}
\put(124.6,19){\line(1,0){30.2}}
\put(143,20){\line(-1,1){6}}
\put(143,20){\line(-1,-1){6}}
\put(161.5,20){\line(1,0){30.2}}
\put(7.5,5){$1$}
\put(44,5){$2$}
\put(82,5){$3$}
\put(119,5){$4$}
\put(156,5){$5$}
\put(193,5){$6$}
\end{picture}
\caption{\label{fig:Borcherds11}\it Dynkin diagram of a Borcherds Lie superalgebra that is related to a quotient of the free Lie superalgebra of $D=11$ supersymmetry.}
\end{figure}

We can also apply a Borcherds construction to the eleven-dimensional case. The starting point for the Dynkin diagram is that of $B_5$ which is the complexification of $\mf{so}(1,10)$. To obtain a spinor representation at level $\ell=1$ one can attach a gray node to the `spinor node' of the $B_5$ diagram with a single line. This leads to figure~\ref{fig:Borcherds11}. The level decomposition of the Borcherds algebra defined by this diagram can be computed using the methods described above and this leads to the $\mf{so}(1,10)$ representation listed in table~\ref{tab:B11}.

\begin{table}[t!]
\centering
\renewcommand{\arraystretch}{1.3}
\begin{tabular}{c|c|c}
$B_\ell$& $B_5$ representation & generator\\
\hline\hline
\multirow{2}{10mm}{\centering$B_0$} & $[0,1,0,0,0]$ & $M_{ab}$\\
& $[0,0,0,0,0]$ & $d$\\\hline
$B_1$ & $[0,0,0,0,1]$ & $Q_\alpha$\\\hline
\multirow{2}{10mm}{\centering$B_2$}  & $[1,0,0,0,0]$ & $P_a$\\
& $[0,1,0,0,0]$ & $P_{ab}$\\\hline
\multirow{2}{10mm}{\centering$B_3$} & $[0,0,0,0,1]$ & $\Sigma_\alpha$\\
 & $[1,0,0,0,1]$ & $\Sigma_{a\alpha}$\\\hline
\multirow{7}{10mm}{\centering$B_4$} & $2\times [0,1,0,0,0]$ & $Z_{ab}=Z_{[ab]}$\\
& $[1,0,0,0,0]$ & \\
& $[0,0,1,0,0]$ & \\
& $[1,1,0,0,0]$ & \\
& $[0,0,0,0,2]$ & \\
& $[1,0,1,0,0]$ & \\
& $[1,0,0,0,0]$ & \\\hline
 $\vdots$& $\vdots$ & $\vdots$
\end{tabular}
\caption{\label{tab:B11}\it The non-negative levels of the Borcherds superalgebra defined by the Dynkin diagram of figure~\ref{fig:Borcherds11}. This is a quotient of the free Lie superalgebra shown in table~\ref{tab:SM11} and the notation is as in that table.}
\end{table}

Table~\ref{tab:B11} should be compared to table~\ref{tab:SM11} from which we see that the Borcherds construction contains considerably fewer generators than the free Lie superalgebra. This is of course expected since the positive levels of the Borcherds algebra are a quotient of the free Lie superalgebra by the Serre ideal. One noteworthy difference between the Borcherds algebra and the free Lie superalgebra is that the former does not contain the generator $P_{a_1\ldots a_5}$ on level $\ell=2$ associated with the M$5$-brane term in the superalgebra. Thus it agrees with
the algebra given in~\cite{Ravera:2018vra} up to $\ell=2$, but it extends it beyond that.\footnote{As in footnote~\ref{fn:black}, we note that it is in fact possible to get also $P_{a_1\ldots a_5}$ at level $\ell=2$ in a Borcherds superalgebra, but then one has to interchange the gray node $6$ in figure~\ref{fig:Borcherds11} for a black node such that the corresponding diagonal entry in the Cartan matrix is not zero but negative. This removes the Serre relation $\{e_6,e_6\}=0$ and gives rise to a 5-form since $\{e_6,e_6\}$ then is a lowest weight vector of the representation $[0,0,0,0,2]$. }

\section{Conclusion}
\label{sec:conc}

In this paper, we have put forward the structure of a free Lie superalgebra for studying extensions of the Poincar\'e superalgebra. The various examples that we have presented comprise generalisations existing in the literature by quotienting and thus the free Lie superalgebra $\mf{f}(Q)$ can be viewed as universal structure that encompasses all these proposals. We have focussed in this paper on the algebraic aspects and have not worked out any dynamical model realising the symmetry but this can be done for the point particle along the lines of~\cite{Bonanos:2009wy,Gomis:2017cmt} and we expect a similar universal behaviour of a superparticle moving in a supersymmetric Maxwell background if a suitable `unfolded' quotient is applied. An aspect that will deserve detailed study in this context is the $\kappa$-symmetry of the particle action. Extensions to string and brane models moving in backgrounds can also be constructed. A generalisation of the Polyakov-type term for the string invariant under the superalgebra can be written as
\begin{align}
\int d^2 \sigma \sqrt{-\gamma} \gamma^{ij} \left[ L_{i}^a L_j^b \eta_{ab} + \frac12 L_i^{ab} L_j^{cd} \eta_{ac} \eta_{bd} + \ldots \right]\,,
\end{align}
where the $L_i^a = \partial_i x^a + \ldots$ and $L_i^{ab} = \partial_i \theta^{ab} + \ldots$ denotes the components of the Cartan--Maurer forms of a string moving on the group manifold of the superalgebra~\cite{Gomis:2017cmt,Bergshoeff:1995hm}. 

Let us also point out a few other possible applications or connections to other work. 
As was emphasised in section~\ref{sec:FLAs} the construction of a free Lie (super-)algebra starts from a set of basic generators, e.g., the $Q_\alpha$. Demanding that these basic generators transform in a representation of some level $\ell=0$ algebra $\mf{g}_0$ then leads to each level $\ell$ of the free Lie algebra $\mf{f}(Q)$ transforming under $\mf{g}_0$. In the discussion of the free Lie superalgebra for $D=11$ in section~\ref{sec:FLA11} we considered the case $\mf{so}(1,10)$, the eleven-dimensional Lorentz algebra. In view of possible relations to $E_{11}$ it is worth noting that the Lorentz algebra is generalised to an algebra called $K(E_{11})$ (or sometimes $I_C(E_{11})$) in the context of West's conjectures. This is an infinite-dimensional Lie algebra that is fixed by an involution acting on (the split real) $E_{11}$, and $K(E_{11})$ contains $\mf{so}(1,10)$. The general representation theory of this algebra has not been classified as it is not of standard type. However, it is known that $K(E_{11})$ can be made to act on the same $32$ supertranslations $Q_\alpha$ that $\mf{so}(1,10)$ acts on~\cite{Kleinschmidt:2006tm,Damour:2006xu}. In this way, the whole free Lie superalgebra $\mf{f}(Q)$ can be seen as a representation of $K(E_{11})$ providing an infinite-dimensional realisation of $K(E_{11})$ albeit still of the same unfaithful type. Under $K(E_{11})$ one has for instance that $\ell=2$ of $\mf{f}(Q)$ forms an irreducible $528$-dimensional representation. This perspective opens up the possibility of constructing particle or brane models with $K(E_{11})$ symmetry.

Yet another possibility of applying the free Lie superalgebra is in the context of so-called vector supersymmetry~\cite{Casalbuoni:2008iy,Barducci:2018wuj}. In vector supersymmetry, the basic supersymmetry generators are not spinors $Q_\alpha$ but transform in a vector and a scalar representation of the Lorentz group.   A particle realisation of this vector supersymmetry leads after quantisation to two uncoupled Dirac equations
~\cite{Casalbuoni:2008iy,Barducci:2018wuj}.

In the physical realisations of Maxwell symmetries studied to date, the background field was non-dynamical. The proposals of~\cite{Borisov:1974bn,West:2001as,Riccioni:2009hi} attempt to also derive the equations of a gravitational background from an algebraic formulation while~\cite{Boulanger:2015mka} discusses the unfolding of the Maxwell equation for the gauge field itself. These constructions are based on algebras that are sometimes called Ogievetskii algebras that include translation generators and higher rank cousins of the type contained in free Lie algebras $\mf{f}(P)$, however, the commutation relations appear to differ from those of the free Lie algebras. It is not clear to us how to reconcile these two pictures at the moment.

\subsection*{Acknowledgements}
We would like to thank Carlo Meneghelli, Hermann Nicolai and Ergin Sezgin for helpful discussions. AK and JP thank the University of Barcelona for its hospitality. JG and AK would also like to thank the University of Brussels and the International Solvay Institutes for hospitality. JG has been supported in part by MINECO FPA2016-76005-C2-1-P and Consolider CPAN and by the Spanish goverment (MINECO/FEDER) under 
 project MDM-2014-0369 of ICCUB (Unidad de Excelencia Mar\'\i a de Maeztu). The work of JP is supported by the Swedish Research Council, project no.~2015-04268.
It began at the Mitchell Institute for Fundamental Physics and Astronomy at the Texas A\&M University and was then supported in part by NSF grant PHY-1214344.

\appendix

\section{Gamma matrices in $D=4$}
\label{gammapp}

An explicit real representation of the $\Gamma^a$ matrices can be written as
\begin{align}
\Gamma^0 &= \begin{pmatrix} 
0 & 0 & 0 &1\\ 
0 & 0 & -1 & 0\\
0 & 1 & 0 & 0 \\
-1 & 0 & 0 & 0
\end{pmatrix}\,\quad&
\Gamma^1 &= \begin{pmatrix} 
0 & 1 & 0 &0\\ 
1 & 0 & 0 & 0\\
0 & 0 & 0 & 1 \\
0 & 0 & 1 & 0
\end{pmatrix}\,\nn\\
\Gamma^2 &= \begin{pmatrix} 
-1 & 0 & 0 &0\\ 
0 & 1 & 0 & 0\\
0 & 0 & -1 & 0 \\
0 & 0 & 0 & 1
\end{pmatrix}\,\quad&
\Gamma^3 &= \begin{pmatrix} 
0 & 0 & 0 &-1\\ 
0 & 0 & 1 & 0\\
0 & 1 & 0 & 0 \\
-1 & 0 & 0 & 0
\end{pmatrix}\,\nn\\
C =  \Gamma^0 &=\begin{pmatrix} 
0 & 0 & 0 &1\\ 
0 & 0 & -1 & 0\\
0 & 1 & 0 & 0 \\
-1 & 0 & 0 & 0
\end{pmatrix}\,\quad&
\Gamma_5 &=\Gamma^0\Gamma^1\Gamma^2\Gamma^3 =\begin{pmatrix} 
0 & -1 & 0 &0\\ 
1 & 0 & 0 & 0\\
0 & 0 & 0 & 1 \\
0 & 0 & -1 & 0
\end{pmatrix}\,.
\end{align}
The matrix $\Gamma^0$ is anti-symmetric while the $\Gamma^i$ (for $i=1,2,3$) are symmetric. They satisfy $\left\{\Gamma^a,\Gamma^b\right\}= 2\eta^{ab}$ with $\eta^{ab} = \textrm{diag}(-1,1,1,1)$. One also has $C^T=-C=C^{-1}$ and that
\begin{align}
\label{eq:gammasym}
(C\Gamma^a)^T = C\Gamma^a \,,\quad (C\Gamma^{ab})^T = C\Gamma^{ab}\,,\quad (C\Gamma^{abc})^T = -C\Gamma^{abc}\,,\quad (C\Gamma^{abcd})^T = - C\Gamma^{abcd}\,,
\end{align}
where $\Gamma^{ab} = \Gamma^{[a} \Gamma^{b]} = \frac12 (\Gamma^a\Gamma^b - \Gamma^b \Gamma^a)$ etc. The matrix $\Gamma_5$ anti-commutes with all the $\Gamma^a$. It squares to $(\Gamma_5)^2= -\mathbb{1}$ and $(C\Gamma_5)^T=- C\Gamma_5$.

We use the following explicit spinor index notation on the gamma matrices, following the SW--NE contraction convention,
\begin{align}
(\Gamma^a)^\alpha{}_\beta \quad\Rightarrow \quad (\Gamma^a \Gamma^b)^\alpha{}_\beta =(\Gamma^a)^\alpha{}_\gamma (\Gamma^b){}^\gamma{}_\beta
\end{align}
together with $C_{\alpha\beta}$ for the anti-symmetric charge conjugation matrix and $C^{\alpha\beta}$ for its inverse satisfying $C_{\alpha\gamma} C^{\gamma\beta} = \delta_\alpha^\beta$. The spinor index range is $\alpha=1,\ldots, 4$. The first relation in~\eqref{eq:gammasym} reads in indices
$(C\Gamma^a)_{\alpha\beta} = (C\Gamma^a)_{\beta\alpha}$, 
where $(C\Gamma^a)_{\alpha\beta} = C_{\alpha\gamma} (\Gamma^a)^\gamma{}_\beta$. We also note the cyclic identity that is satisfied by the gamma matrices in $D=4$ are
\begin{subequations}
\label{eq:cyclics}
\begin{align}
\label{eq:cycl}
(C\Gamma_a)_{(\alpha\beta} (C\Gamma^a)_{\gamma)\delta} = 0 \,,
\\\label{eq:cycl2}
(C\Gamma_{ab})_{(\alpha\beta}(C\Gamma^{ab})_{\gamma)\delta}=0\,,
\\\label{eq:cycl3}
(C\Gamma^a)_{(\alpha\beta}C_{\gamma)\delta} + (C\Gamma^{ab})_{(\alpha\beta} (C\Gamma_b)_{\gamma)\delta}=0\,.
\end{align}
\end{subequations}
The identity (\ref{eq:cycl}) and similar ones in $D=3,6,10$ are important for the construction of supersymmetric Yang--Mills theories~\cite{Brink:1976bc,Gliozzi:1976qd}, whereas (\ref{eq:cycl2}) is crucial for supersymmetric membranes~\cite{Achucarro:1988qb}.  In the context of free Lie superalgebras they appear in the Jacobi identity~\eqref{eq:SJ}.

\end{document}